\PassOptionsToPackage{table,dvipsnames,xcdraw}{xcolor} 

\documentclass[a4paper]{article}

\usepackage[english]{babel}
\usepackage[utf8x]{inputenc}
\usepackage[T1]{fontenc}

\usepackage[a4paper,top=3cm,bottom=2cm,left=3cm,right=3cm,marginparwidth=1.75cm]{geometry}

\usepackage{amsmath}
\usepackage{amsfonts} % for Real number symbol etc
\usepackage{bm} % for bold math symbols
\usepackage{graphicx}
\usepackage[colorlinks=true, allcolors=blue]{hyperref}

\usepackage{xcolor} 
\usepackage{parskip} 
\usepackage{cleveref} 
\usepackage{enumitem} 
\usepackage[framemethod=TikZ]{mdframed} % Create box
\usepackage{authblk} % for author affiliations 

\newcommand{\indep}{\perp \!\!\! \perp}

\title{A tutorial on individualized treatment effect prediction from randomized trials with a binary endpoint}

\author[1]{J Hoogland}
\author[2]{J IntHout}
\author[2]{M Belias}
\author[2]{MM Rovers}
\author[3]{RD Riley}
\author[4]{FE Harrell Jr}
\author[1,5]{KGM Moons}
\author[1,5]{TPA Debray}
\author[1,5]{JB Reitsma}

\affil[1]{Julius Center for health sciences and primary care, University Medical Center Utrecht, Utrecht University, Utrecht, the Netherlands}
\affil[2]{Radboud Institute for Health Sciences (RIHS), Radboud university medical center, Nijmegen, The Netherlands}
\affil[3]{School of Medicine, Keele University, Keele, Staffordshire, UK}
\affil[4]{Department of Biostatistics, Vanderbilt University School of Medicine, Nashville TN, USA}
\affil[5]{Cochrane Netherlands, University Medical Center Utrecht, Utrecht University, Utrecht, the Netherlands}
\date{}                     
\begin{document}

%%TC:ignore

\maketitle

\begin{abstract}
Randomized trials typically estimate average relative treatment effects, but decisions on the benefit of a treatment are possibly better informed by more individualized predictions of the absolute treatment effect. In case of a binary outcome, these predictions of absolute individualized treatment effect require knowledge of the individual’s risk without treatment and incorporation of a possibly differential treatment effect (i.e. varying with patient characteristics). In this paper we lay out the causal structure of individualized treatment effect in terms of potential outcomes and describe the required assumptions that underlie a causal interpretation of its prediction. Subsequently, we describe regression models and model estimation techniques that can be used to move from average to more individualized treatment effect predictions. We focus mainly on logistic regression-based methods that are both well-known and naturally provide the required probabilistic estimates. We incorporate key components from both causal inference and prediction research to arrive at individualized treatment effect predictions. While the separate components are well known, their successful amalgamation is very much an ongoing field of research. We cut the problem down to its essentials in the setting of a randomized trial, discuss the importance of a clear definition of the estimand of interest, provide insight into the required assumptions, and give guidance with respect to modeling and estimation options. Simulated data illustrates the potential of different modeling options across scenarios that vary both average treatment effect and treatment effect heterogeneity. Two applied examples illustrate individualized treatment effect prediction in randomized trial data. 

\end{abstract}

%TC:endignore

\section{Introduction}

Prediction of risk and prediction of treatment effect are two key components in modern medicine and personalized healthcare. On the one hand, risk predictions are classically functions of multiple patient characteristics. They include predictions of the risk of having a specific health outcome or condition (diagnosis) or of developing a  future health outcome (prognosis). Also, risk predictions vary naturally across patients, are descriptive, and can be uniformly expressed as probabilities \cite{steyerberg_clinical_2009}. Importantly, risk prediction models are generally descriptive and are \textit{not} intended to reflect the causal mechanism; in particular, included predictor effects in the model are not intended to reflect the extent to which their removal or modification would change an individual’s prediction. On the other hand, predictions of treatment effect \textit{do} express an expected difference due to modification of the treatment condition. They have classically been studied on a group level (e.g. treated group versus control group), often assume a constant effect across individuals, have a causal interpretation, and are traditionally expressed using relative effect measures (\textit{e.g.} odds ratio, relative risk, or hazard ratio) \cite{senn_statistical_2007}. 

Risk predictions and treatment effect estimation are two important areas of research but have largely developed in separation, leading to an apparent contradiction between methods for prediction and methods for causal inference. However, answers to many important questions need to bridge the divide. For instance, "How will a possible treatment change predicted outcome risk?" or "Is there variability in the effect of this treatment across patients (\textit{i.e.} differential treatment effect)?". These questions involve both the causal effect of treatment on a targeted health outcome and the adequate incorporation of associations with individual patient characteristics.  

It is exactly these types of questions that need to be answered to provide more tailored, stratified, personalized, or precision medicine \cite{glasziou_evidence_1995, kent_personalized_2018, kent_predictive_2020}. The limitations of average relative treatment effects have long been recognized and the promise of a more individualized yet evidence-based approach has been enticing. Such a strategy requires focus on heterogeneity between patients and its relation to risk of the outcome of interest and variability in treatment effect. Also, moving towards more individualized estimates inherently means moving to more absolute expressions of variability in risk and treatment effect that are interpretable on the individual level \cite{dahabreh_using_2016, murray_patients_2018}. For example, predictions of the absolute risk of a future event under different treatment conditions provide a natural basis for shared decision-making. 

In this tutorial, we aim to give a platform for statisticians and other researchers embarking on the prediction of individualized treatment effect. We focus on the risk of developing a binary outcome or endpoint, and aim to combine the highly conditional nature of typical risk prediction modeling with causal inference about treatment effectiveness. While the problem is well-known, it is very complex and therefore, we will cut it down to its essentials in the setting of a randomized trial, and mainly limit our scope to regression-based methods. We discuss the importance of a clear definition of the estimand of interest, provide insight into the required assumptions, and give guidance with respect to the modeling and estimation options. Key considerations with respect to the choice of modeling and estimation methods are further illustrated in a simulation study and two applied examples.

\section{Defining individualized treatment effect} \label{sec:delta_i}

The main idea underlying our endeavor is that the effect of treatment may be different for each individual, and that it may be beneficial to personalize or individualize its estimate. In the context of risk prediction, this implies that treatment causes a change in predicted outcome risk that may vary across individuals conditional on their characteristics. In other words, a personalized or individualized treatment effect describes the effect of modifying a treatment condition (\textit{i.e.} setting its value) while controlling for (\textit{i.e.} conditioning on) that individual's characteristics. We restrict our description to settings in which variables besides treatment do not have a causal interpretation, since this nicely aligns with the typical design of a randomized trial. This lack of causal interpretation for the set of variables conditioned on, is typical for classical prediction modeling. While the inner workings of a model that simultaneously describes both causal and mere associative relations may not need to discern between these different roles, they are of importance when interpreting the model. 

To that effect, distinguishing between variables that do and do not have a causal interpretation is helpful for a precise definition of the individualized treatment effect of interest. Two common approaches to make this distinction are the \textit{do($\cdot$)} operator introduced by Pearl \cite{pearl_causal_2016} and the potential outcomes framework popularized by Rubin \cite{rubin_causal_2005}. The \textit{do($\cdot$)} operator is an operator that describes the effect of setting or modifying a variable to take a certain value (e.g. $P(Y=y|do(X=x))$) and clearly separates this case from classical conditional notation (e.g. $P(Y=y|X=x)$). The potential outcomes framework, as popularized by Rubin, allows for a formal distinction at the level of the outcomes that arises when the causal variable takes on different values \cite{rubin_estimating_1974, rubin_causal_2005}. For instance, if interest is in the causal effect of a treatment, and treatment takes values $a \in \mathcal{A}$, $Y^{A=a}$ denotes the potential outcome for treatment $a$. In case of a treatment variable that can be set to 0 (control) or 1 (treated), the two potential outcomes are $Y^{a=0}$ and $Y^{a=1}$. The notation easily allows for conditioning, such that the effect $\delta$ of treatment on the risk of an event, conditional on covariates $\bm{X}$, can be written as

\begin{equation} \label{eq:deltaX}
  \delta(\bm{x}) = P(Y^{a=1}=1|\bm{X}=\bm{x}) - P(Y^{a=0}=1|\bm{X}=\bm{x})
\end{equation}

where bold face indicates vectors. The same quantity could be written in \textit{do($\cdot$)} notation as 

\begin{equation}\label{eq:deltaXdo}
  \delta(\bm{x}) = P(Y=1|do(A=1),\bm{X}=\bm{x}) - P(Y=1|do(A=0),\bm{X}=\bm{x})
\end{equation}

For our purposes, the differences between these frameworks are not of interest and we adopt the potential outcomes framework throughout the remainder of the paper for reasons of familiarity in statistical research.

A final remark on the nature of 'individualized' or 'personalized' is in place: the estimand of interest is not truly individual, but relates to groups of individuals sharing a covariate pattern. A truly individual treatment effect can never be observed since only one potential outcome can be observed at any time (or equivalently, only one treatment can be assigned). This problem has been referred to as the fundamental problem of causal inference \cite{holland_statistics_1986}. Acknowledging this, for a dichotomous outcome $Y$, we define the individualized treatment effect ($\delta(\bm{x}_i)$) for individual $i$ with covariate vector $\bm{x_i}$ as 

\begin{equation} \label{eq:delta_i}
  \delta(\bm{x}_i) = P(Y_i^{a=1}=1|\bm{X}=\bm{x_i}) - P(Y_i^{a=0}=1|\bm{X}=\bm{x_i})
\end{equation}

The individualized treatment effect $\delta(\bm{x}_i)$ can be interpreted as the expected difference in outcome risk for an individual with covariate values $\bm{x}_i$ under two different treatment conditions. This definition is easily extended to $\geq 2$ treatment conditions, but we will focus on a setting with 2 treatment conditions. While we will focus on $\delta(\bm{x}_i)$ as our estimand of interest, note that important information is lost when only looking at the difference between potential outcomes. Therefore, the main task is to predict the conditional risk of both potential outcomes, which provides a more complete picture and directly leads to an estimate of $\delta(\bm{x}_i)$. Section \ref{sec:identifiability} first discusses the assumptions that allow estimation of $\delta(\bm{x}_i)$ from randomized trial data. Subsequently, sections \ref{sec:models} and \ref{sec:estimation} describe specification and estimation of regression models for the purpose of predicting $\delta(\bm{x}_i)$.

\section{Identifiability assumptions} \label{sec:identifiability}
The individualized treatment effect specified in equation \eqref{eq:delta_i} is written as a difference between two potential outcomes. However, in practice only a single potential outcome will be observed for each individual. Identification of $\delta(\bm{x}_i)$ based on the observed data requires assumptions to supplement the data. The necessary identifiability assumptions are consistency, exchangeability, and positivity. We here shortly introduce the fundamentals as relevant to our setting; excellent introductory \cite{hernan_definition_2004} and comprehensive texts on causal inference are available elsewhere \cite{hernan_causal_2020}. 

\textit{Consistency} refers to equality between the observed outcome $Y$ and the potential outcome for the actually assigned treatment $Y^a$. When $A$ takes on value $0$ (control) or $1$ (treated), this can be expressed as $Y = AY^{a=1} + (1-A)Y^{a=0}$. This assumption holds when the data reflect well-defined treatments. As a counter example, consider a situation in which there is much variability in the active treatment (e.g. different starting times, intensity or dosage, duration) but these are all just labelled $a=1$: the causal contrast $Y^{a=1} - Y^{a=0}$ is no longer clearly defined. As is clear from equation \eqref{eq:delta_i}, the contrast of interest actually requires consistency conditional on covariates $\bm{X}$. This extension is trivial if marginal consistency holds. 

\textit{Exchangeability} requires that the potential outcomes are independent of treatment assignment ($Y^a \indep A$ for all $a$). In other words, the actually assigned treatment does not predict the potential outcome \cite{hernan_definition_2004}. As an example, consider a two-arm study of a new treatment: in terms of potential outcomes, exchangeability with respect to treatment here implies that it does not matter which arm received the new treatment. Since our interest is in a conditional treatment effect, exchangeability should hold conditionally ($Y^a  \indep A|\bm{X}$ for all $a$). While this is a challenging assumption to satisfy in general, it holds automatically in the context of a randomized trial. After either marginal randomization (\textit{i.e.} a common probability of treatment for all) or conditional randomization (\textit{i.e.} with the probability of treatment depending on covariates, also known as stratified randomization), conditional exchangeability holds when conditioning on (at least) the variables used during randomization. It is important to realize that randomization only provides exchangeability at baseline, and the causal contrast $\delta(\bm{x}_i)$ at that time therefore reflects an intention-to-treat effect. Any conditioning on post-randomization information is no longer protected by randomization and the exchangeability assumption will no longer be guaranteed to hold \cite{hernan_randomized_2013, hernan_per-protocol_2017}. For instance, estimation of a per protocol individualized treatment effect would require further assumptions such as absence of any unmeasured confounders and correct specification of all confounders \cite{goetghebeur_formulating_2020}. We here limit our overview to intention-to-treat effects.

\textit{Positivity} reflects the assumption that each patient should have a non-zero probability of either treatment assignment, which is clearly fulfilled in case of a randomized study. 

A final assumption that is often made is the assumption of \textit{no interference}, stating that the potential outcomes for one individual do not depend on treatment assignment of other individuals. While not strictly necessary, the situation quickly grows in complexity without this assumption since the potential outcome definitions would then have to incorporate the dependence on other units \cite{rubin_causal_2005, goetghebeur_formulating_2020}. The combination of consistency and no interference is also often referred to as the \textit{stable unit treatment value assumption} (SUTVA) \cite{rubin_causal_2005}. 

The definition of $\delta(\bm{x}_i)$ in equation \eqref{eq:delta_i} assumes no interference, which follows from the fact that the potential outcome for individual $i$ only depends on the individual's own covariate status and treatment assignment. Further assuming positivity provides a causal interpretation of the treatment effect conditional on covariates $\bm{x}_i$, with $i$ from $1,\ldots,n$. Finally, consistency and exchangeability are necessary to re-write the estimand in terms of observed variables only:
\begin{align} 
\delta(\bm{x}_i) &= P(Y_i^{a=1}=1|\bm{X}=\bm{x_i}) - P(Y_i^{a=0}=1|\bm{X}=\bm{x_i}) \nonumber \\ 
                 &= P(Y_i^{a=1}=1| A=1, \bm{X}=\bm{x_i}) - P(Y_i^{a=0}=1| A=0, \bm{X}=\bm{x_i}) \quad \textnormal{(by exchangeability)} \nonumber \\
                 &= P(Y_i=1| A=1, \bm{X}=\bm{x_i}) - P(Y_i=1| A=0, \bm{X}=\bm{x_i})  \quad \textnormal{(by consistency)} \label{eq:delta_iobs}
\end{align}

In summary, the identifiability assumptions allow $\delta(\bm{x}_i)$ to be estimated from the observed data. While equation \eqref{eq:delta_iobs} essentially allows for fully non-parametric estimation of $\delta(\bm{x}_i)$, there is usually insufficient data to do so when interest is in a highly conditional treatment effect (as is the case for an individualized treatment effect). That is, there will not be sufficient cases with $\bm{X}=\bm{x}_i$ under both treatments to reliably estimate $\delta(\bm{x}_i)$. This brings us to the need of a model for $P(Y_i=1|A=a_i,\bm{X}=\bm{x}_i)$ to smooth over the gaps in the observed set of all $\bm{x}_i$ across both treatments. 

\section{Models for the prediction of individualized treatment effect} \label{sec:models}

With the identifiability conditions in place for a causal interpretation of $\delta(\bm{x}_i)$ as estimated based on the observed trial data only, the remaining problem can be recognized as a typical prediction modeling problem (equation \eqref{eq:delta_iobs}). Therefore, well-established modeling techniques can be used to model the required conditional risks. While a vast array of possible prediction modeling techniques is available, we will focus on modeling techniques that have a basis in generalized linear modeling. More specifically, due to the binary outcome, we will focus on methods that have a basis in logistic regression, which has been the mainstay method for clinical prediction models in settings with a binary endpoint  \cite{steyerberg_clinical_2009}. Key features of logistic regression include that it directly provides the probabilistic estimates of interest \cite{spiegelhalter_probabilistic_1986} and has well-known properties. Also, it is a possibly parsimonious model family for the task at hand as explained in the next section. 

\subsection{Homogeneous treatment effect} \label{sec:constantTE}

\begin{figure}[!htb]
\centering
\includegraphics[width=0.6\linewidth]{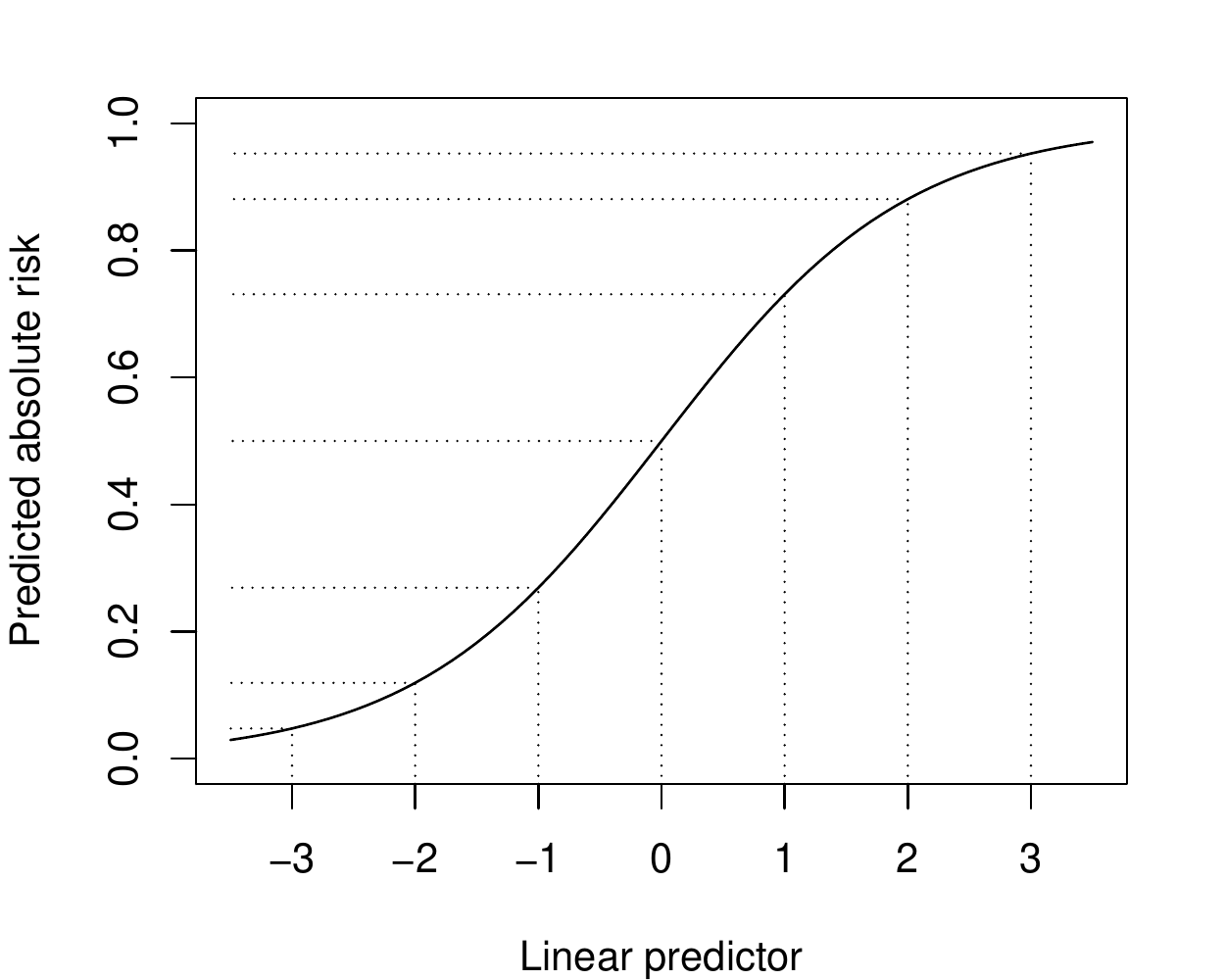}
\caption{The translations of effects from the linear predictor (LP) scale to the absolute scale vary depending on location as shown for the logistic link function ($\frac{1}{1+e^{-\textnormal{LP}}}$). Therefore, a constant or homogeneous relative effect, here shown as a 1-point difference on the linear predictor (\textit{i.e.} log odds) scale for ease of exposition, has different implications on the absolute risk scale. For example, for a patient with a control risk of 50\% (LP=$0$), a treatment effect of -1 on the log odds scale reduces predicted absolute risk to 27\% (LP=$-1$, resulting in an absolute risk reduction of  $23\%$). For a patient with a control risk of $27\%$ (LP=$-1$), the same treatment effect on the log odds scale leads to a predicted risk of $12\%$ (LP=$-2$, resulting in an absolute risk reduction of $15\%$).}
\label{fig:location}
\end{figure}

The aim is to model the observed outcomes $\bm{Y}$ as a function of treatment assignment $A$ and covariates $\bm{X}$ to provide estimates for the right-hand side of equation \eqref{eq:delta_iobs} and hence $\delta(\bm{x}_i)$. While the treatment effect of interest is expressed in terms of a difference in probabilities when the outcome is binary, this may not be the most appropriate scale to model treatment effect. The reason that the logistic model might provide a parsimonious model to predict the required conditional probabilities, is that a constant effect on the log odds scale has a valid interpretation across the entire range of predicted probabilities. For instance, the effect of treatment on outcome risk could really be constant on the log odds scale regardless of outcome risk in absence of treatment. This property does not hold for linear probability models or relative risk models, but very similar results may be obtained for probit models. As a quick reminder, Figure \ref{fig:location} shows the logistic link that transforms the log odds or linear predictor scale to the probability scale. A constant treatment effect on the log odds scale has a large effect on the probability scale when the linear predictor equals zero and approaches zero for very low and high linear predictor values. This nicely reflects the difference in the amount of wiggle room when the control outcome risk reflects unpredictability versus near certainty respectively. The reason to emphasize these well-known properties is to highlight that a very simple model on the log odds scale may very well lead to potentially relevant differences between individuals on the level of $\delta(\bm{x}_i)$ (\textit{i.e.} differences in absolute risk). 

\vspace{5pt}

The simplest way to create such a model is to assume absence of any interaction between treatment and the other covariates. Thus, the (conditional) treatment effect is assumed to be \textit{homogeneous} or constant on the log odds scale. In terms of notation, let $Y_{i}$ denote the independent dichotomous \textit{observed} outcomes, assumed to have patient specific mean $\mu_i$ (\textit{i.e.} $\smash{Y_{i} \overset{i.i.d.}{\sim} \textnormal{Bernoulli}(\mu_{i})}$). Also, with vector $\bm{x}_i$ denoting the $p$ individual characteristics and $a_i$ being the treatment indicator as before, this leads to the following simple logistic regression model 

\begin{align} \label{eq:logistic_additive}
\begin{split}
\textnormal{logit}(P(Y_{i} = 1 | A=a_i, \bm{X}=\bm{x}_{i})) =  
    \beta_0 + \beta_t a_i + \bm{\beta}^\top \bm{x}_{i}
\end{split}
\end{align}

with regression parameters $\beta_0$, $\beta_t$, and $\bm{\beta} = (\beta_1, \ldots, \beta_p)$. The key assumptions for this logistic model including a conditional \textit{homogeneous} treatment effect $\beta_t$ are i) appropriateness of the logistic link function, ii) linearity in the parameters, and iii) additivity of at least the treatment effect on the log odds scale (\textit{i.e.} there are no treatment-covariate interactions on the log odds scale). The linearity assumption on the covariate contributions could easily be relaxed, allowing for global or local transformations of $\bm{x}_{i}$ such as polynomials and splines. 

The predicted individualized treatment effect follows directly after estimation of the model parameters \footnote{For our current goal and scope, inferring individualized treatment effect from predictions under the relevant potential outcomes is the end-point. However, it is interesting to note that this type of conditional potential outcome predictions can also serve as input for substitution estimators aiming for more marginal estimates, such as the parametric g-formula \cite{hernan_causal_2020} and estimators of marginal risk difference and marginal risk ratio based on logistic models \cite{austin_absolute_2010}.}: 
\begin{align} \label{eq:delta_ihat}
\begin{split}
\hat{\delta}(\bm{x_i}) = \frac{1}{1 + e^{-(\hat{\eta}_i + \hat{\beta}_t)}} - \frac{1}{1 + e^{-\hat{\eta}_i}}
\end{split}
\end{align}
where $\hat{\eta}_i = \hat{\beta}_0 + \bm{x}_{i}^\top \hat{\bm{\beta}}$. Strict additivity of the treatment effect on the log odds scale may provide a parsimonious model for the analysis and is easily translated to the more interpretable scale of ${\delta}(\bm{x_i})$ where this additivity does not hold. Note that in case of such a homogeneous treatment effect on the log odds scale, variability in $\bm{x}_{i}$ is the driving force behind any differentiation on the level of $\hat{\delta}(\bm{x_i})$. This variability in $\bm{x}_{i}$ corresponds to variability in prognosis across individuals under control treatment (\textit{i.e.} variability in $Y_i^{a=0}$).

\subsection{Heterogeneous or differential treatment effect} \label{sec:hte}
As an extension of homogeneous treatment effect models, the relative treatment effect can also be allowed to depend on the other covariates. We will refer to such non-additivity of the treatment effect on the relative scale as heterogeneity of treatment effect (HTE) or differential treatment effect. We note  that there is no single accepted definition of the term HTE in the literature, and that it is sometimes used in a broader sense to also include the variability in ${\delta}(\bm{x_i})$ that may result from a homogeneous treatment effect model \cite{kent_personalized_2018}. We use HTE in its narrow sense (\textit{i.e.} restricted to non-additive relative effects), since this allows one to distinguish between possible variability in the way treatment affects different individuals (homogeneous versus heterogeneous) and variability amongst individuals that does not relate to treatment effect (\textit{i.e.} variability in expected prognosis under control treatment $P(Y_i=1| A=0, \bm{X}=\bm{x_i})$).

The homogeneous treatment effect model in equation \eqref{eq:logistic_additive} can easily be extended to allow for HTE by inclusion of treatment-covariate interactions. The model then becomes 
\begin{align} \label{eq:logistic_hte}
\begin{split}
\textnormal{logit}(P(Y_{i} = 1 | A=a_i, \bm{X}=\bm{x}_{i})) = 
    \beta_0 + \beta_t a_i + \bm{\beta_m}^\top \bm{x}_{i}  + 
    \bm{\beta_z}^\top \bm{z}_{i} a_i
\end{split}
\end{align}
where $\bm{z}_{i}$ is a subset of $\bm{x}_{i}$, $\bm{\beta_{m}}$ includes the coefficients for the main effects of $\bm{x}_{i}$, and $\bm{\beta_{z}}$ includes the coefficients for treatment-covariate interactions. As before, the space of the measured $\bm{x}_{i}$ can be expanded using global or local transformations. These transformations need not be the same in $\bm{z}_{i}$: the functional form of the effect of a covariate may depend on treatment status. Also, note that when $\bm{z}_{i}$ equals $\bm{x}_{i}$ (\textit{i.e.} all covariates are involved in treatment-covariate interactions), an exactly equivalent parametrization can be obtained by specification of separate models for the treated group and the control group with just an intercept and main covariate effects, which separates the models for both potential outcomes. Additional details are provided in the online supporting material (Part \ref{app:perArm}). 

Based on the model in equation \eqref{eq:logistic_hte}, the predicted individualized treatment effect again follows easily from the parameter estimates. In analogy to equation \eqref{eq:delta_ihat}, the prediction of ${\delta}(\bm{x_i})$ can be derived according to 
\begin{align} \label{eq:delta_ihat_hte}
\begin{split}
\hat{\delta}(\bm{x_i}) = \frac{1}{1 + e^{-(\hat{\eta}_i + \hat{\beta}_t + 
\bm{\hat{\beta}_z}^\top \bm{z}_{i})}} - \frac{1}{1 + e^{-\hat{\eta}_i}}
\end{split}
\end{align}

While this prediction comes nice and easy in theory, the challenge lies in precise model specification and the estimation of the model parameters of these relatively complex models.

\section{Model estimation} \label{sec:estimation}

The preferred method for estimation of the prediction models of interest depends on the relation between model complexity and the amount of signal in the data. In medical statistics, the amount of variability in the outcome of interest that can be explained is often low to moderate, which is the main reason for the need for large sample sizes. In case of insufficient sample size, models are prone to overfitting, which describes the situation where the model captures part of the noise in the data. Overfitting is an important concern since it limits generalizability of the model. In this section we discuss the need for methods that mitigate the susceptibility to overfitting.

\subsection{Maximum likelihood}

Estimates for the proposed logistic regression models can be derived with standard maximum likelihood estimation for generalized linear models \cite{nelder_generalized_1972, agresti_foundations_2015}. However, even after taking all content knowledge into account, models are often still overly complex with respect to the available sample size. This may already hold for a logistic model of homogeneous treatment effect when the number of covariates is large, or their functional form allowed to be complex and the sample size is relatively small \cite{riley_calculating_2020}. It has long been recognized that standard maximum likelihood estimation of logistic models is problematic in these settings due to finite sample bias, perfect separation, collinearity, and overfitting \cite{spiegelhalter_probabilistic_1986, van_smeden_sample_2018}. Models including many non-additive effects such as treatment-covariates interactions are even more prone and necessitate either strong prior assumptions or restrictions during model estimation.

\subsection{Penalized maximum likelihood} \label{sec:pML}
Regression based on penalized maximum likelihood estimation (pML) has emerged as a method that can, at least to some degree, cope with relatively complex models \cite{hoerl_ridge_1970, tibshirani_regression_1996, efron_least_2004}. These methods penalize the log-likelihood for the magnitude of regression coefficients other than the intercept. This penalty introduces bias towards zero on the estimated (non-intercept) coefficients, and thus towards the overall outcome incidence for predictions. In other words, it introduces a bias that reduces variability at the level of the predictions. This balance is also known as the bias-variance trade-off. Well-known penalized maximum likelihood methods include ridge regression, lasso regression, and the elastic net which includes the first two as special cases \cite{friedman_regularization_2010}. The ridge penalty is a smooth penalty on squared size of the regression parameters and leads to shrinkage of the estimated coefficients. It was originally developed to deal with collinearity and tends to distribute the weight amongst collinear variables \cite{hoerl_ridge_1970}. The lasso penalty is on the absolute value of the coefficients and leads to both shrinkage and selection \cite{tibshirani_regression_1996}. It has a tendency to select amongst collinear variables. The elastic net penalty is a weighted balance of the ridge and lasso penalty. The required degree of penalization is a tuning parameter that needs to be estimated from the data, which is most commonly done by means of cross-validation. Importantly, this estimation involves uncertainty that is most problematic when accurate penalization is needed the most (\textit{i.e.} small data sets and/or low signal relative to noise) \cite{van_calster_regression_2020, riley_penalisation_2020}. With respect to the equivalence of a model including all treatment-covariate interactions on the one hand and separate modeling in the treated and control group on the other hand (section \ref{sec:hte}), note that this equivalence no longer holds in case of penalized maximum likelihood. Details are provided in the online supporting material (Part \ref{app:perArm}).

\subsubsection{Shrinkage and/or selection}
\textbf{Penalization in clinical prediction modeling}
In some settings, the underlying process to be modelled is fairly well know, and therefore, the same holds for the elements that should be included in a model. While such a setting does not require selection of parameters, shrinkage may still be beneficial in terms of prediction accuracy when sample size is limited with respect to model complexity. 
In contrast, the available data may be very rich while the underlying process not well understood, but thought to be sparse. Penalization approaches that provide selection (\textit{i.e.} sparse solutions) have been successfully applied to prediction problems with many covariates (possibly more than cases, \textit{i.e.} $p>>n$) where selection of variables is key and there is insufficient content knowledge to do so, such as the selection of possibly important signals from microarray data \cite{hastie_statistical_2015}. 

In the typical context of clinical prediction modeling, the properties of the problem are somewhere in-between these two extremes. In the best-case scenario, clear pre-specification of a model might be possible. Often however, even though the data are typically low-dimensional, some further variable selection may still be required. The choice of penalty, and hence the need for selection, therefore heavily depends on the state of content knowledge and the amount of data available. An issue that may guide the selection of a penalty function is the need for an honest representation of the relative weights of model parameters. Ridge regression tends to share the regression weights between parameters that are correlated with the outcome \textit{and} with each other \cite{hastie_ridge_2020}, while lasso tends to select amongst such variables \cite[Chapter~4]{hastie_statistical_2015}. As an example, if two highly correlated covariates are equally predictive of the outcome, ridge will keep both in the model with approximately equal weight. Lasso will remove the one variable that happens to have a slightly weaker association with the outcome in the current sample. Both representations can be useful, but they serve a different purpose. 

\textbf{Penalization and modeling of heterogeneous treatment effect}
Models of heterogeneous treatment effect include treatment-covariate interactions. Such interactions are always harder to estimate than the overall (main) treatment effect. In terms of selection, heterogeneous treatment effect models encounter the variable selection issue twice: for the main effects and for the treatment-covariate interactions. Usually, lower order (main) effects are kept in the model for each component of an interaction. However, both lasso regression and elastic net regression do not respect this hierarchical nature. To that effect, a hierarchical group lasso algorithm has been developed that does respect the hierarchy between main effects and interaction effects \cite{lim_learning_2015}. In short, variable selection is achieved on a group level that is allowed to be hierarchical, such that main effect groups will be in the model when they are part of any interaction. For problems with non-overlapping groups, regular group lasso algorithms are also widely available \cite{yang_fast_2015}.

\section{Model complexity}

The current state of subject matter or content knowledge, the type of process under study, the strength of the associations in the data, and the final purpose of the model, all weigh into the decision on the best balance between prior model specification and more data-driven modeling methods. We have described both specification and estimation of logistic models that can be used to predict treatment effect. In this section, we will shortly touch upon the complementary roles of content knowledge and penalization, the concern of overfitting when aiming for out-of-sample prediction, and the degree to which model complexity can be left to the data-drive methods.

\subsection{Content knowledge and penalization}
Content knowledge, general statistical knowledge, and data-driven methods such as penalization should work synergistically to arrive at the most parsimonious model that reflects current content knowledge as updated by the data. Ideally, content knowledge includes knowledge on non-linear covariate contributions, interactions amongst covariates, and especially knowledge on covariates that might interact with treatment. Unfortunately, such knowledge is often very limited. Consequently, the number of parameters that one wants to estimate may still be relatively large even after all content knowledge is exhausted. While data driven approaches can help to nudge such models in the right direction, they provide no panacea or substitution of content knowledge. 

\subsection{Overfitting and prediction accuracy}
For our aim to predict individualized treatment effect, the conditional model of treatment effect is intended to generalize and therefore overfitting is a key concern. The primary challenge is to balance model complexity with respect to the amount of available data in a way that generalizes beyond the original sample. The preferred way to do so is to limit model complexity based on content knowledge, supplemented by use of penalization. The relation between prediction accuracy, model complexity, effective sample size, and the strength of the associations in the data is well known and has been described in the context of risk prediction (\textit{e.g.} \cite{van_smeden_sample_2018, riley_calculating_2020}). However, estimation of $\delta(\bm{x}_i)$ requires the difference between two outcome risk predictions (under the two to be compared treatments) to be accurate. While these two predictions will be highly correlated since they arise from the same individual, the expected error will invariably be larger than for a single prediction. To our knowledge, there is no guidance available on the necessary conditions with respect to effective sample size and expected explained variation for accurate prediction of risk differences within individuals. Our simulation study below provides some first insights. 

\subsection{'Risk modeling' versus 'effect modeling'} 
The use of treatment-covariate interactions to model heterogeneous treatment effect (\textit{e.g.} as in equation \eqref{eq:logistic_hte}) has also been referred to as 'effect modeling' and has been distinguished from 'risk modeling' \cite{kent_personalized_2018, kent_predictive_2020, sussman_improving_2015, kent_assessing_2010, burke_using_2014}. In risk modeling, treatment effect variability is evaluated as a function of outcome risk, where outcome risk is a function of the covariates. Therefore, it can be seen as a data reduction method. For example, in risk modeling the linear predictor scores $\hat{\eta_i}$ resulting from a model for $P(Y_i=1| A=0, \bm{X}=\bm{x_i})$ can be interacted with treatment instead of the full multi-dimensional representation of $\bm{x_i}$. That is,
\begin{align} \label{eq:riskModeling}
\textnormal{logit}(P(Y_{i} = 1 | A=a_i, \hat{\eta_i}) = \beta_t a_i + \hat{\eta_i} + f(\hat{\eta_i}) a_i
\end{align}
where $f(\cdot)$ denotes a possibly flexible function and $\hat{\eta_i}$ is used as an offset. The general idea is that $f(\hat{\eta_i})$ is a much simpler structure to estimate than many individual treatment-covariate interactions. Therefore, it is supposed to fill a gap in situations where content knowledge is insufficient to limit model complexity to something that can be reliably estimated in the data.

From a statistical point of view, 'risk modeling' does of course reduce the risk of overfitting, since it restricts the modeled treatment effect heterogeneity to be a function of a scalar. However, the price to pay is that HTE is thereby forced to be proportional to the main effects of $\bm{x_i}$. This implies i) that all covariates that have a main effect also modify the relative treatment effect, and ii) that the effect of each element of $\bm{x_i}$ on HTE has the same direction as its effect on outcome risk. These are strong assumptions that have no clear biological substrate and no clear statistical preference over other data reduction methods such as principal component analysis \footnote{The preference to model HTE as a function of outcome risk was originally motivated as "outcome risk is a mathematical determinant of treatment effect" \cite{kent_personalized_2018}, along with a reference to the fact that the [marginal] odds ratio [of exposure to treatment] equals $EER / (1-EER) \div CER / (1-CER)$ [where EER is the experimental outcome prevalence and CER is the control outcome prevalence]. However, the same equation can be used to show that independence is a possibility: i) only the $CER / (1-CER)$ part depends on control outcome risk and is positive and finite for any control prevalence in the $(0,1)$ interval, and ii) combined with any valid odds ratio, $EER / (1-EER)$ is also positive and finite and therefore maps back to a prevalence amongst the treated in the $(0,1)$ interval. This also holds for the conditional treatment effect and was in fact one of the reasons to prefer logistic regression as explained in section \ref{sec:constantTE}.}. Nonetheless, the idea behind risk modeling does reflect recognition of the danger of overfitting when modeling many treatment-covariate interactions, which remains an important issue when modeling heterogeneous treatment effect. 

\subsection{Tree-based methods} 
Many different modeling techniques are available under the machine learning umbrella. While we have limited our scope to regression-based methods, there are other methods that could be used instead. In particular, several tree-based methods have been developed for the specific purpose of individualized treatment effect prediction. While an in-depth discussion is beyond our scope, we here provide several key references. Wager et al. \cite{wager_estimation_2018} extend the well-known random forest algorithm by Breiman \cite{breiman_random_2001} to enable individualized treatment effect prediction and provide a very thorough overview of the required conditions for causal inference, including asymptotic theory. Also, they provide an overview of the literature on forest-based algorithms for estimating heterogeneous treatment effect. Lu et al. \cite{lu_estimating_2018} provide a clear exposition of individualized treatment effect prediction and the potential outcome framework, and provide empirical and simulation results on a wide variety of random forest methods for causal inference, including virtual twins, counterfactual random forests, the aforementioned causal forests and Bayesian adaptive regression trees. Generalized random forests \cite{athey_generalized_2019} constitute a more recent addition to the literature and form a much broader method that can also be used for causal individualized treatment effect estimation (and effectively encompass the work by Wager et al. \cite{wager_estimation_2018}). Lastly, model-based recursive partitioning is a somewhat different tree-based approach that incorporates parametric models into the tree \cite{zeileis_model-based_2008}. Such a parametric model can for instance describe control outcome risk and relative treatment effect. Node-splitting then occurs on the variable that generates most instability in the parameters for this model. Seibold et al. \cite{seibold_model-based_2016} have developed a model-based recursive partitioning random forest to identify treatment effect heterogeneity. Model-based recursive partitioning is most closely related to the methods discussed in this tutorial since it can differentiate between heterogeneity of treatment effect on the relative scale and differential outcome risk that may be related to a homogeneous treatment effect (i.e. constant log odds of treatment). 

Beyond the predictions of individualized treatment effect, there have been efforts to find subgroups that might need different treatment based on a fitted causal forest \cite{scarpa_assessment_2019}, and efforts to further explain random forest-based treatment effect predictions based on covariate data \cite{lu_estimating_2018}.  Both papers go through considerable lengths to disentangle and interpret predicted individualized treatment effect differences. While this may lead to interesting hypotheses, this should be a careful undertaking. An illustration of the things that could go wrong is available in Rigdon et al. \cite{rigdon_preventing_2018}, showing high false discovery rates if such care is not taken. 

To the best of our knowledge, direct comparisons of regression-based and other methods for the prediction of individualized treatment effect have not been performed yet. In general, it can be expected that more flexible models are more prone to overfitting and require more data \cite{van_der_ploeg_modern_2014}. A challenge in the comparison of different methods in simulation studies is that specific data generating mechanisms favor specific methods. For instance, a data generating mechanism with linear and additive effects will favor regression methods; a mechanism generating subgroups based on cut-offs will favor tree-based algorithms. An interesting method to acknowledge these effects and check robustness is provided by Austin et al \cite{austin_predictive_2021}, and could also be applied in the context of individualized treatment effect prediction in future work comparing a wider set of methods.

\newcounter{theo}\setcounter{theo}{0}
\renewcommand{\thetheo}{\arabic{theo}}
\newenvironment{theo}[2][]{%
\refstepcounter{theo}%
\ifstrempty{#1}%
{\mdfsetup{%
frametitle={%
\tikz[baseline=(current bounding box.east),outer sep=0pt]
\node[anchor=east,rectangle,fill=black!20]
{\strut Box~\thetheo};}}
}%
{\mdfsetup{%
frametitle={%
\tikz[baseline=(current bounding box.east),outer sep=0pt]
\node[anchor=east,rectangle,fill=black!20]
{\strut Box~\thetheo:~#1};}}%
}%
\mdfsetup{innertopmargin=10pt,linecolor=black!20,%
linewidth=2pt,topline=true,%
frametitleaboveskip=\dimexpr-\ht\strutbox\relax
}
\begin{mdframed}[]\relax%
\label{#2}}{\end{mdframed}}

\section{Learning from simulations} \label{sec:sim}

We conducted a simulation study to illustrate the consequences of the choice of model and estimation method when predicting individualized treatment effect. Such simulations are especially helpful when predicting potential outcomes, since they can never be observed directly in practice. Also, the ability to manipulate the data generating mechanism into several interesting settings has a clear illustrative advantage. In the design of our simulation study, we adhered to the general guidelines proposed by Burton et al. \cite{burton_design_2006} and Morris et al. \cite{morris_using_2019}. 

\subsection{Data generating mechanisms} \label{sec:DGMs}
The data generating mechanism was parametric and was based on a logistic model (equation \eqref{eq:logistic_hte}). Settings varied across the full factorial combination of varying total sample size (400, 1200, 3600), presence/absence of a main treatment effect ($\beta_t=ln(0.6)$ or $\beta_t=ln(1)$), and presence/absence of heterogeneity of treatment effect. The treatment indicators $a_i$ were independent samples from a Bernoulli distribution with probability $0.5$. Twelve covariates were drawn from a multivariate standard normal distribution with a compound symmetric covariance matrix ($\rho = 0.1$). Main effect coefficients $\bm{\beta_m}$ were of exponentially decreasing size ($\beta_{m,1}, \ldots, \beta_{m,12}=2^{-\frac{0}{2}}, 2^{-\frac{1}{2}}, 2^{-\frac{2}{2}}, \ldots, 2^{-\frac{11}{2}}$) to reflect i) decreasing added value of consecutive variables, and ii) that it is unlikely for variables that are included in a risk model to have truly zero coefficients. In settings with a homogeneous treatment effect, there were no treatment-covariate interactions (\textit{i.e. $\bm{\beta_z}=\bm{0}$}). In settings with a heterogeneous treatment effect, $\bm{\beta_z}$ was equal to $-\frac{1}{2},-\frac{1}{4}$ and $-\frac{1}{8}$ for $\beta_{z,10}, \beta_{z,11}$ and $\beta_{z,12}$ respectively, and included a small random perturbation that was generated once for all simulations ($\leq |0.05|$) for $\beta_{z,1}, \ldots, \beta_{z,9}$. In each simulation setting, the intercept was chosen such that the true underlying outcome prevalence in the control arm was 25\% (details provided in the online supporting material, Part \ref{app:eventRate}). Nagelkerke $\textnormal{R}^2$, as measured between the true conditional probabilities for the assigned treatment (\textit{i.e.} $P(Y_i^{A=a_i}=1 | \bm{X} = \bm{x}_i$) and the observed events $Y_i$ in a large sample, was approximately 0.4 for all settings.

\begin{figure}[!htb]
\centering 
\includegraphics[width=.9\linewidth]{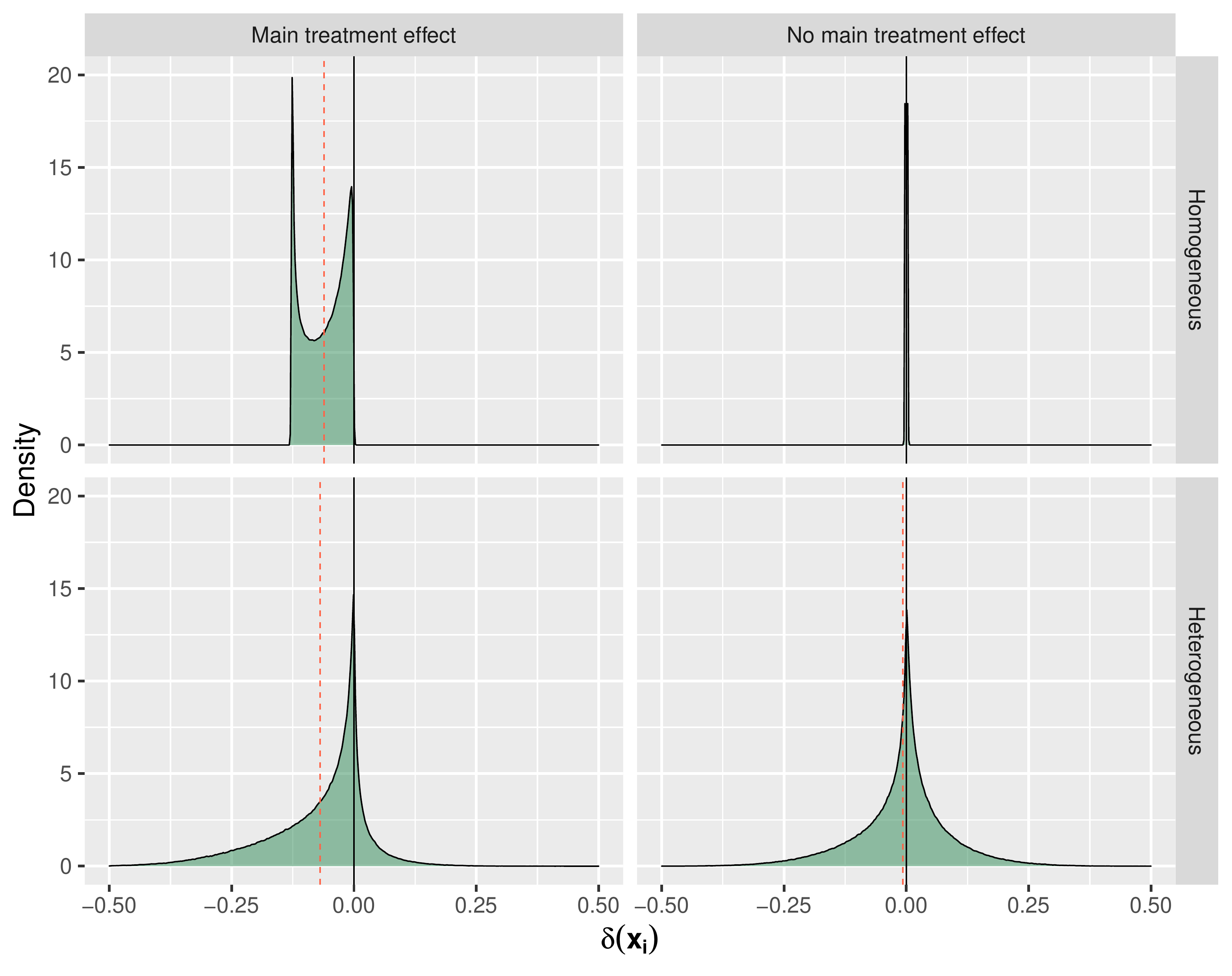}
\caption{Distribution of $\delta(\bm{x}_i)$ according to the data generating mechanism for each of the simulation settings. The quadrants correspond to settings with a homogeneous main treatment effect (upper left), absence of any treatment effect (upper right), heterogeneous treatment effect in presence of a main treatment effect (lower left), and heterogeneous treatment effect in absence of a main treatment effect (lower right). The red dotted lines provide the mean of $\delta(\bm{x}_i)$ per setting. Note that all of the mass in the upper right figure is on a spike at $\delta(\bm{x}_i) = 0$.}
\label{fig:sim_delta_i}
\end{figure}

To get more insight into the different simulation settings, the distribution of $\delta(\bm{x}_i)$ for each of the data generating mechanisms is shown in Figure \ref{fig:sim_delta_i}. Note that i) there is no variability at all in the upper right figure (due to absence of any treatment effect); ii) variability in the upper left figure is due to variability in control risk across patients, and iii) variability in the lower two figures is due to heterogeneity in both control risk and treatment effect. 

\subsection{Model development} \label{sim:modeldevelopment}
Within each simulation run, both a development and a validation data set were simulated for each of the simulation settings. That is, both data sets were always generated according to the same data generating mechanism. The size of the model development sets matched the simulation settings (\textit{i.e.} $400, 1200$ or $3600$), and the size of the validation sets was always equal to 10,000 observations.

Table \ref{tab:simulationApproaches} provides an overview of the evaluated methods. The overall absolute treatment effect is just the marginal version of $\delta(\bm{x}_i)$ as marginalized over all covariates. Its estimate of $\delta(\bm{x}_i)$ is just the $\hat\delta$, the difference in mean outcome incidence between treatment arms. For the homogeneous treatment effect models, all covariates entered the model only as main effects. In case of heterogeneous treatment effect models, all covariates entered the model as both main effects and interactions with treatment. The selected penalty parameter for ridge, lasso, and hierarchical group lasso (HGL) was the one with the smallest deviance in 10-fold cross-validation. Note that both lasso and HGL may set coefficients to exactly zero (\textit{i.e.} perform selection). Also, while HGL can search the entire interaction space, the current implementation was limited to treatment-covariate interactions in parallel to the other methods. A final variation of the HTE methods was a 'content knowledge' (CK) setting, where we assumed that content knowledge suggests that only the first eight variables have important main effects, and only covariates nine to twelve are likely treatment interaction candidates. Ridge regression was used to estimate such a CK-based model. The 'risk modeling' implementation included a risk model estimated in the control group based on main effects for all covariates and a linear treatment with risk-score interaction. Both standard maximum likelihood and ridge regression were performed for the risk model. A significance-based approach was implemented as a final comparison. Starting from a homogeneous treatment effect model including all covariates, a likelihood ratio test was performed for the treatment effect coefficient. When non-significant, the treatment coefficient was removed from the model (leaving only main covariate effects). When significant, all treatment-covariate interactions were added to the model and a second likelihood ratio test was performed to evaluate their joint significance. Treatment-covariate interactions were kept in the model when this joint test was significant and removed otherwise. All tests used an $\alpha$ level of $0.05$. In addition to the methods in Table \ref{tab:simulationApproaches}, HTE was modeled using separate prediction models per treatment arm, and thus per potential outcome, as described in the online supporting material (Part \ref{app:perArm}).

\begin{table}[!ht]
\centering
\begin{tabular}{lcr}
\hline
\bf{Regression model} & \bf{Equations} & \bf{Estimation method} \\
\hline
\hline
Overall absolute treatment effect (Overall) & --  & ML \\

Homogeneous treatment effect (HOM)  & \eqref{eq:logistic_additive} \eqref{eq:delta_ihat}  & ML, ridge \\

Heterogeneous treatment effect (HTE, HTE-CK)*  & \eqref{eq:logistic_hte} \eqref{eq:delta_ihat_hte} & ML, $\textnormal{ridge}^{\dagger}$, lasso, HGL \\

'Risk modeling' (RM) &\eqref{eq:riskModeling} & ML, ridge \\

Significance-based (SB) & -- & ML \\

\hline
\end{tabular}
\caption{\label{tab:simulationApproaches} Implemented methods towards the prediction of individualized treatment effect ($\hat{\delta}(\bm{x}_i)$). 
\newline * Model specification differs between the default case (HTE) modeling all main effects and treatment-covariate interactions, and the content knowledge case (HTE-CK) modeling a selection of main effects and an treatment-covariate effects (details are described in Section \ref{sim:modeldevelopment}).
\newline $\dagger$ Only ridge was used for the HTE-CK model. 
\newline Abbreviations: ML (Maximum Likelihood), HGL (Hierarchical group lasso).}
\end{table}

\subsection{Model evaluation} \label{sec:performance_measures}
Each of the methods provides a prediction vector $\bm{\hat{\delta}}$ with elements $\hat{\delta}_i$ (short for $\hat{\delta}(\bm{x}_i)$) as derived in the validation sample. These predictions can be compared to the known $\bm{\delta}$ based on the data generating mechanism. The root mean squared prediction error (rMSPE) between the elements of these two vectors was used to quantify the prediction errors according to

\begin{equation} \label{eq:rmspe}
\textnormal{rMSPE} = \sqrt{n^{-1} (\bm{\hat{\delta}}-\bm{\delta})^\top (\bm{\hat{\delta}}-\bm{\delta})}.    
\end{equation}

The root was taken to arrive at an expression of the error on the risk difference scale. In addition, the 0.9-quantile of absolute prediction errors was derived for both $\bm{\hat{\delta}}$ and predicted risk. Supplementing these single figure summaries, calibration plots were derived for both $\bm{\hat{\delta}}$ and predicted risk, where predictions were cut into twenty equal-size quantiles groups and compared to the true values based on the data-generating mechanism. 

\subsection{Statistical software} \label{sec:implementation}
All analyses were performed in {\tt R} statistical software version 3.5.0 \cite{r_core_team_r:_2017}. The \texttt{R} script for exact replication of the simulation study is available for sharing. Logistic regression models based on maximum likelihood estimation were fitted using {\tt glm()}. Ridge and lasso implementations were based on the {\tt glmnet} package \cite{friedman_regularization_2010}. Hierarchical group lasso was implemented using the {\tt glinternet} package \cite{lim_learning_2015}.

\subsection{Simulation study results} \label{sec:sim results}
We here synthesize the results of 250 simulation runs in terms of root mean squared prediction error and calibration of the predicted individualized treatment effect.

\subsubsection{Average root mean squared prediction error}
\begin{figure}[!htb]
\centering
\includegraphics[width=.9\linewidth]{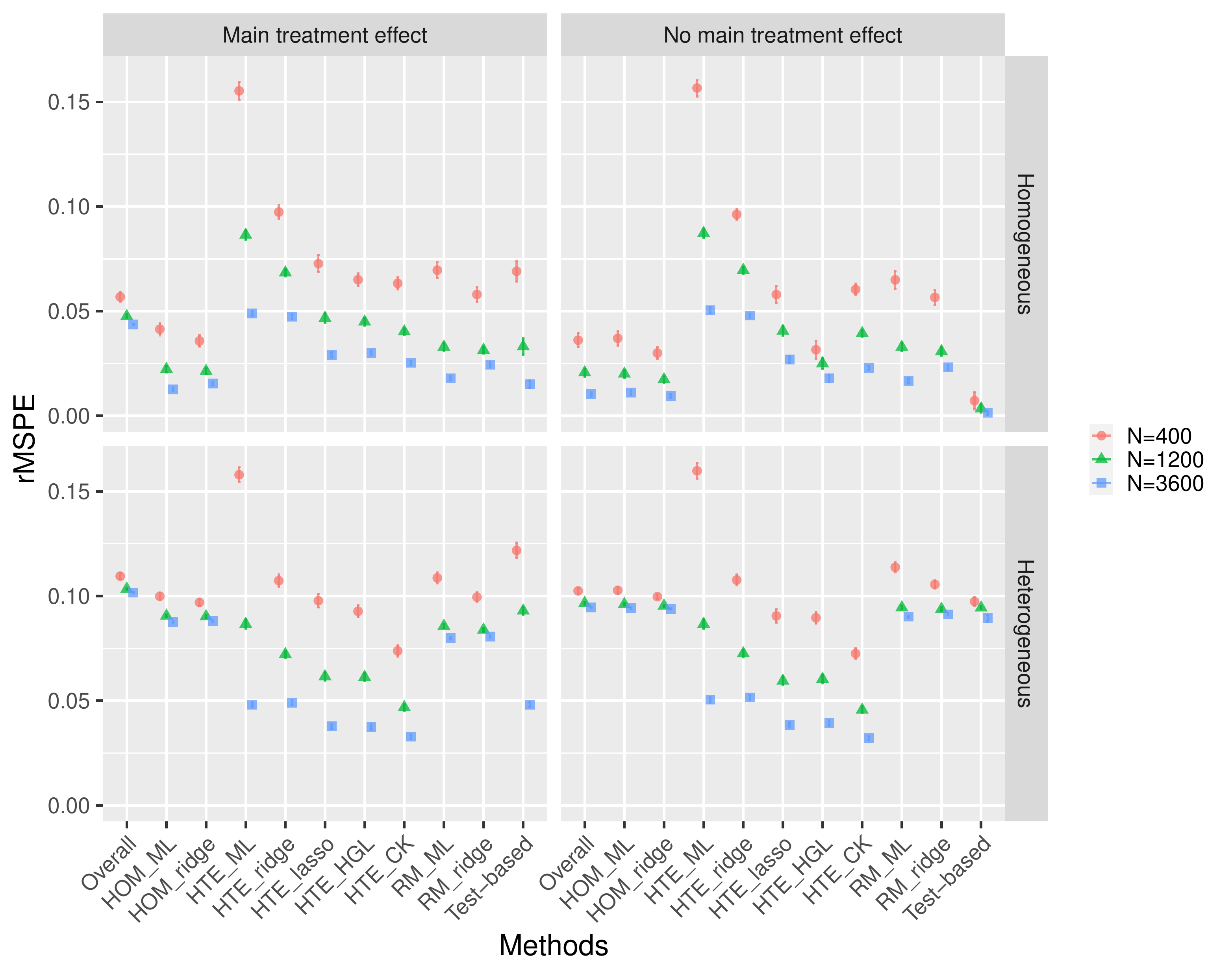}
\caption{Simulation study results: average root mean squared prediction error (rMSPE) of the predicted treatment effects (over 250 simulations) with $\pm 2$ SE error bars for all simulation settings. The quadrants correspond to settings with a  homogeneous main treatment effect (upper left), absence of any treatment effect (upper right), heterogeneous treatment effect in presence of a main treatment effect (lower left), and heterogeneous treatment effect in absence of a main treatment effect (lower right). Note that the standard errors are often so small that they are obscured by the mean estimates. Abbreviations: homogeneous treatment effect models (HOM), heterogeneous treatment effect models with treatment-covariate interactions (HTE), and heterogeneous treatment effect models based on risk modeling (RM), as estimated by means of maximum likelihood (ML), ridge or lasso regression, or hierarchical group lasso (HGL), with CK standing for content knowledge.}
\label{fig:rmspe}
\end{figure}

The main simulation results with respect to root mean squared prediction error (rMSPE) are shown in Figure \ref{fig:rmspe}, which provides a summary of the error that can be expected in the long run across settings and methods. Five key observations can be made across all settings. First, conditioning on main effects of the available covariates as in a homogeneous treatment effect model was always beneficial when compared to the fully marginal $\hat{\delta}$. This confirms the idea that a conditional estimand ($\delta(\bm{x}_i)$) requires a conditional estimator. Second, HTE model accuracy was especially sensitive to sample size, which is in line with expectation due to the amount of parameters that needs to be estimated. Third, penalization is key and improved estimation of both homogeneous and heterogeneous treatment effect models up to large sample sizes. Fourth, penalization does \textit{not} remove the risk of overfitting for complex models in small sample size settings. Heterogeneous treatment effect models could not be reliably fitted in small samples regardless of the estimation method. Fifth, utilization of content knowledge was the most effective way to reduce HTE model complexity. Lasso and HGL did not catch up, even though the content knowledge simulated here was not entirely correct and the lasso models did start from the correct set of variables given the data generating mechanism. The reduction in the potential set of treatment-covariate interaction variables in the content knowledge-based model was very effective, even though this missed out on small interaction effects. Nonetheless, apparent content knowledge could of course not help out when it was wrong, such as in the complete null setting (no main treatment effect, no heterogeneity). Lastly, risk modeling was not the preferred method in any of the simulation settings. This of course depends on the data generating mechanism where HTE was not fully explained by a risk model, but this would also not be expected in practice and serves the purpose of showing that risk modeling may miss important treatment effect heterogeneity. Additional rMSPE results for comparing treatment-covariate interaction modeling with separate prediction models per treatment arm, and thus per potential outcome, are described in the online supporting material (Part \ref{app:perArm}). In short, treatment-covariate interaction modeling by means of lasso regression was best across all settings when compared to per arm modeling. The differences between treatment-covariate interaction modeling and per arm modeling were more nuanced for ridge regression.

\subsubsection{Calibration}
The online supplementary material (Part \ref{app:deltaihat_calibration}) provides calibration plots of $\hat{\delta}(\bm{x}_i)$ across methods and settings. It provides further insight into the distribution of the errors and the degree of variability across replications. Several important observations can be made that supplement the conclusions based on rMSPE. First, calibration curves at least pass the $(0,0)$ point, and the size of the errors increases as predictions move away from zero. Therefore, prediction errors were much smaller around the harm-benefit boundary (\textit{i.e.} $\delta(\bm{x}_i)=0$). Second, calibration in individual small samples could be far off even if average performance across simulations was good. For instance, fitting a ridge regression homogeneous treatment effect model in accordance with the data generating mechanism could still lead to substantial overfitting or underfitting in any individual data set. These findings on the risk difference scale are in line with earlier results on direct prediction \cite{van_calster_regression_2020, riley_penalisation_2020}. Third, in small and even medium sample sizes, even penalized heterogeneous treatment effect models overfitted to such an extent that they falsely predicted harm for a subpopulation when in fact there was none (as in the homogeneous treatment effect settings). Fourth, when the data generating mechanism was heterogeneous, calibration of predicted treatment effect was quite reasonable for penalized HTE models in medium and large sample sizes. Note that, while useful for illustrative purposes in this simulation setting, these calibration plots are not available in practice since they compare predictions against the true individual treatment benefits (which do not even have an observed individual level equivalent).

\section{Applied examples} \label{sec:applied_example)}

\subsection{Acute otitis media} \label{sec:aom}
For illustrative purposes, we analyzed data from a randomized, double-blind, placebo-controlled trial of amoxicillin for clinically diagnosed acute otitis media (AOM) in children 6 months to 5 years of age \cite{le_saux_randomized_2005}. This trial included 512 children and collected baseline data on antibiotic treatment received, sex, presence of recurrent AOM, fever, bilateral occurrence, ear pain, presence of a runny nose, cough, tympanic membrane abnormality, and age. All variables but the latter were dichotomous. The endpoint analyzed here is the same as reported by Rovers et al. \cite{rovers_antibiotics_2006}: positive when either fever or ear pain was present after 3 days of follow-up. While not truly binary, composite endpoints occur frequently in practice. Thus, data were available on a total of 9 patient characteristics, treatment, and on a composite dichotomous endpoint. All in all, there were 147 events.

We first fitted a logistic regression model on the full data set with main effects for treatment and the 9 patient characteristics. The estimated log odds of treatment was statistically insignificant, but in the expected direction ($\hat{\beta}_t = -0.34$, se=$0.20$). The apparent Nagelkerke $\textnormal{R}^2$ for this model was only 8.8\%, and a larger sample size would be generally be required for prediction model development in such low signal settings \cite{riley_minimum_2019}. Considering the simulation study results, the sample size, and the low amount of signal with respect to predicted risk, the starting point for any prediction of individualized treatment effect in these data is very weak. Nonetheless, it is interesting to see whether the proposed methods indeed show a lack of predictive ability. To that effect, we internally validated all of the methods evaluated in the simulation study (except for the use of content knowledge) in 100 bootstrap samples. The lowest out-of-sample Brier score (0.202) was obtained with the homogeneous treatment effect model fitted by means of ridge regression. However, the accompanying out-of-sample Nagelkerke $\textnormal{R}^2$ was near-zero and even negative for more flexible models. Together, these findings show a lack of strong support for a non-zero average treatment effect and for any ability to personalize treatment effect.

\subsection{International Stroke Trial} \label{sec:ist}
The International Stroke Trial (IST) was a large randomized open trial comparing no antithrombotic treatment, aspirin treatment, and subcutaneous heparin treatment in a total of 19,435 patients with acute ischaemic stroke \cite{international_stroke_trial_collaborative_group_international_1997}. The individual patient data from the IST trial are available for public use and can be downloaded from the web \cite{international_stroke_trial_collaborative_group_international_2011}. For the current applied example, we used data from patients randomized between no treatment ($n=4,860$) and aspirin treatment ($n=4,858$). Interestingly, the effect of aspirin treatment on the combined endpoint of death or dependency at 6 months was evaluated conditional on a prognostic score in the original article (and found to be effective on average). The prognostic score was based on age, sex, state of consciousness, and 8 other neurological symptoms evaluated at baseline. All variables except for age (continuous) and sex (dichotomous) are categorical variables with three levels. The total number of events was 6,043. Ninety-nine patients with a missing outcome were omitted; covariates were complete. 

We first fitted a logistic regression model on the full data set with a main effect for treatment and main effects for the sex, state of consciousness and the eight  neurological signs. Categorical variables were dummy coded. The estimated log odds of treatment in this model was $-0.11$ (se $0.048$), corresponding to an odds ratio of 0.90, and the apparent C-statistic and Nagelkerke $\textnormal{R}^2$ were 0.79 and 0.31 respectively. Even though the relative treatment effect is small, the presence of an average effect, the ability to explain a substantial part of the risk of an event, and the amount of available data provide a good starting position when aiming to individualize treatment effect prediction. In terms of the simulation study results, the expectancy is that a HTE model would pick up heterogeneity if it is present and that a homogeneous model would be preferred otherwise. 

\begin{figure}[!htb]
\centering
\includegraphics[width=.9\linewidth]{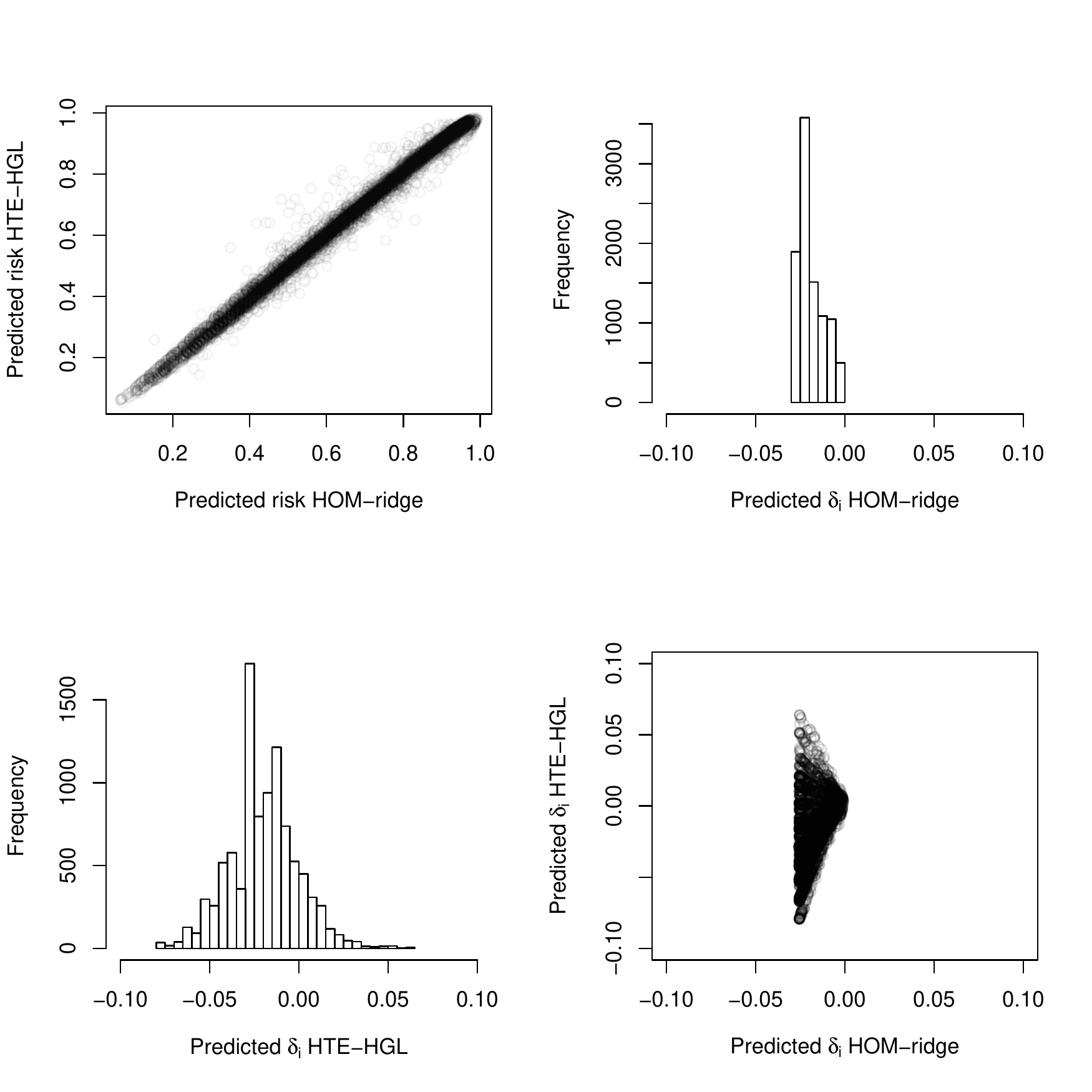}
\caption{International Stroke Trial (IST) data. The upper left panel shows the relation between predicted risks based on the ridge homogeneous treatment effect model (HOM-ridge) and the hierarchical group lasso HTE (HTE-HGL). The histograms show the distribution of predicted individualized treatment effect for the same models. The lower right panel shows their mutual relation.}
\label{fig:ist}
\end{figure}

For illustrative purposes, we therefore examine the predictions from a ridge homogeneous treatment effect model (HOM-ridge) and a hierarchical group lasso HTE (HGL-HTE) model as fitted in the entire sample. In both cases, the penalty parameter with the lowest 10-fold cross-validation deviance was selected. Figure \ref{fig:ist} shows the very high correlation (0.996) between risks predicted from either model (upper left panel). Nonetheless, the distribution of $\hat\delta(\bm{x}_i)$ is quite different for both models (upper right and lower left panel). The HOM-ridge model hardly discriminates w.r.t. treatment effect, whereas the HTE-HGL model predicts harm for a substantial part of the population. The lower right panel shows the relation between $\hat\delta(\bm{x}_i)$ as predicted from both models. 

\begin{figure}[!htb]
\centering
\includegraphics[width=.9\linewidth]{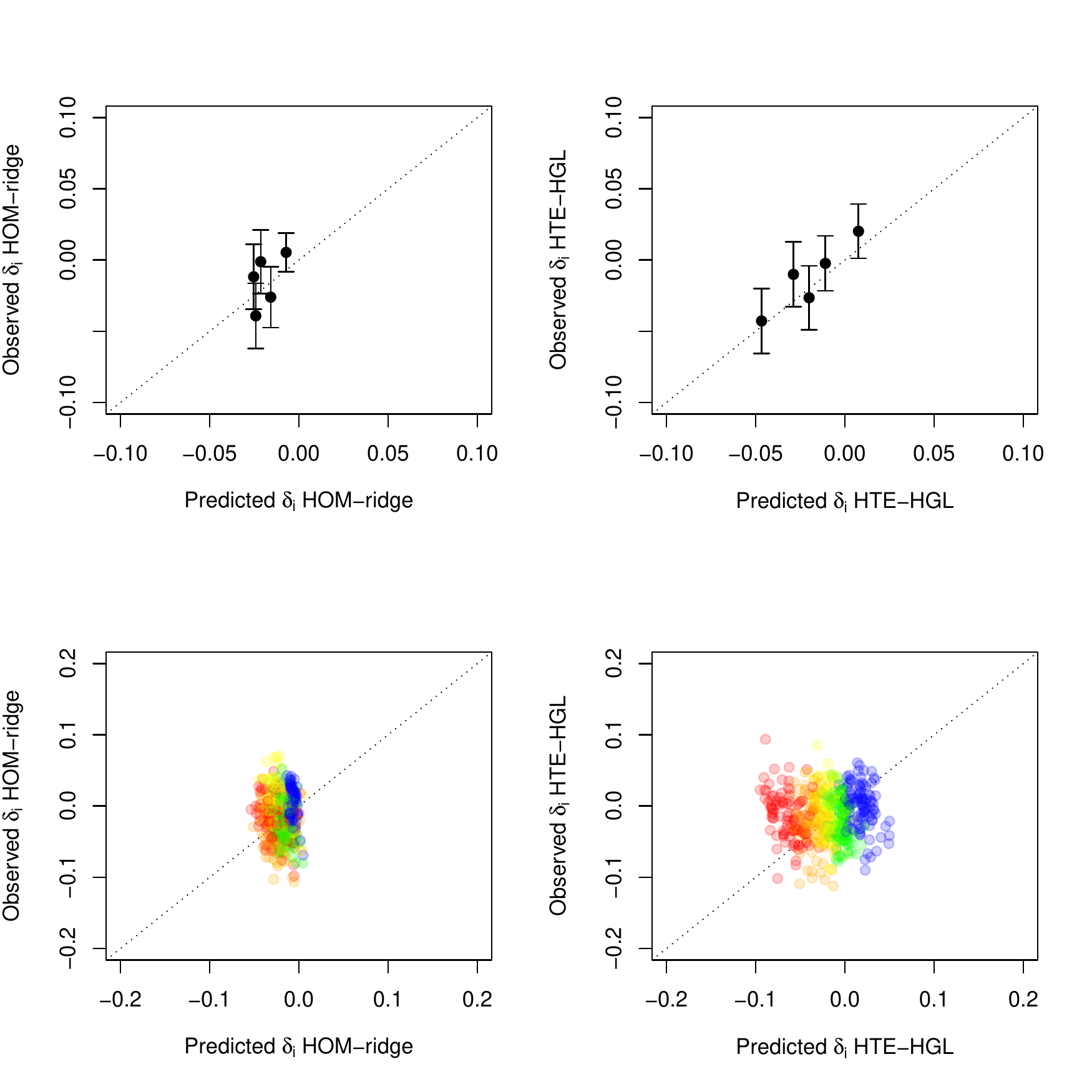}
\caption{International Stroke Trial (IST) data. The upper panels show apparent calibration of treatment effect for the ridge homogeneous treatment effect model (HOM-ridge) and the hierarchical group lasso HTE (HTE-HGL). Dots describe mean $\hat\delta(\bm{x}_i)$ within quintiles of $\hat\delta(\bm{x}_i)$ (x-axis) versus the marginal treatment effect within those same groups (y-axis). Bars provide $\pm1$ SE. The lower panels show the out-of-sample estimates obtained from 100 bootstrap replications with red representing the lowest quintile $\hat\delta(\bm{x}_i)$, followed by orange, yellow, green,  and blue for the highest quintile of $\hat\delta(\bm{x}_i)$.}
\label{fig:ist2}
\end{figure}

The main question is whether any, and if so, which of the predictions of $\delta(\bm{x}_i)$ can be expected to generalize. A key limitation is that $\hat\delta(\bm{x}_i)$ cannot be validated directly in real data. A possible approximation is to check whether groups based on quantiles of $\hat\delta(\bm{x}_i)$ relate to observed treatment effect within these same groups. This is essentially a group level effort to approximate calibration at the level of $\hat\delta(\bm{x}_i)$. The idea is to make use of the fact that the potential outcomes, and thus $\delta(\bm{x}_i)$, are independent of treatment assignment (exchangeability). Figure  \ref{fig:ist2} shows apparent and bootstrap results. As can be seen, the apparent calibration of predicted $\hat\delta(\bm{x}_i)$ is not good for HOM-ridge and reasonable for HTE-HGL. However, in both cases, bootstrap results show such a high degree of variation and lack of trend that the prediction of $\hat\delta(\bm{x}_i)$ cannot be trusted. In case of the HOM-ridge model, this may relate to the fact that such small variations in $\hat\delta(\bm{x}_i)$ cannot just not be retrieved from limited out-of-sample cases. In case of the HTE-HGL model, it implies that even in such a large sample, penalized models may still overfit. 

To further illustrate this, we again applied all methods evaluated in the main simulation study (except for the application of content knowledge). We evaluated model fit by means of out-of-sample Brier score and Nagelkerke $\textnormal{R}^2$ in 100 bootstrap replications. The best Brier score was achieved for the homogeneous treatment effect model fitted with standard maximum likelihood, closely followed by the equivalent ridge version and the hierarchical group lasso (HGL) and ridge HTE model (0.17910, 0.17915, 0.17921, and 0.17933 respectively). The same group of methods had the highest Nagelkerke $\textnormal{R}^2$ values (0.3055, 0.3059, 0.3062, and 0.3066 respectively), with little difference between them. All in all, the differences in bootstrap-corrected estimates of overall fit were very small even though the number of model parameters differed substantially across models. Such results, and the extremely high correlation between risk predictions from different models (as in the left upper panel of Figure \ref{fig:ist}), may already be interpreted as red flags with respect to overfitting of the HTE models. 

\section{Practical considerations} \label{sec:pc}
In general, as with regular prediction modeling, overfitting is an important concern and sample size and penalization are key. We refer to recent guidance papers on sample size \cite{van_smeden_sample_2018, riley_minimum_2019, riley_calculating_2020} and the use of penalization in clinical prediction modeling \cite{van_calster_regression_2020, riley_penalisation_2020}. These guidelines provide a lower bound for the sample size when developing models to predict individualized treatment effects. The required sample size will further increase when including treatment-covariate interactions.

With respect to the choice of modeling approach, the hierarchical group lasso (HGL) performed well across settings. HGL best captured treatment effect heterogeneity when present and had acceptable overfitting otherwise. It clearly outperformed HTE models based on ridge, lasso, or unpenalized maximum likelihood. In practice, a penalized homogeneous treatment effect model provides an important alternative model that is less prone to overfitting, but also less capable of capturing HTE. This modelling approach therefore appears particularly promising when RCTs are relatively small, or when there is (almost) no treatment-effect modification. Cross-validation or bootstrapping may be used to choose between HGL and homogeneous treatment effect models. The International Stroke Trial applied example provided an example of such a bootstrapping approach, showing the evaluation of well-known measures of overall fit (Brier score and Nagelkerke $\textnormal{R}^2$) \cite{steyerberg_clinical_2009, harrell_regression_2015}, and a proposal to visually evaluate performance on the level of aggregated individualized treatment effect.

Nonetheless, samples that are very small with respect to the number of model parameters of interest should raise caution \cite{van_smeden_sample_2018, riley_minimum_2019, riley_calculating_2020}. While penalization is clearly beneficial on average, accurate estimation of the penalization parameter itself cannot be expected in small sample size settings and there is no way to know whether this affects any particular model in practice \cite{van_calster_regression_2020, riley_penalisation_2020}. Also, the empirical evaluation of individualized treatment effect models is still in its infancy. Both group level evaluation of predicted individualized treatment effects and individual level evaluation of outcome risk predictions are only indirect approximations of the performance of individualized treatment effect prediction. Due to the insensitivity of the available measures, we expect model comparisons to be conservative in the sense that they will only prefer a heterogeneous model if the evidence is quite strong. For instance, HTE needs to be large enough to substantially affect overall outcome predictions, or sample size needs to be large enough to reliably assess aggregate predicted versus observed treatment effect. Also, complex models suffer more from the decreased variability in bootstrap samples.

Importantly, any measure based on observed data can only be an approximation of the performance of a potential outcome prediction model, with the remainder resting on assumptions \cite{dickerman_counterfactual_2020}. The plausibility of the identifiability assumptions is an important part of model evaluation, including thoughts about their transportability, and requires thinking instead of measuring. To the knowledge of the authors, there is no specific guidance on the validation of the potential outcome type of individualized treatment effect prediction models beyond the guidance provided in this tutorial. This remains an important open area of research that needs further study.

Lastly, a note on the interpretation of any identified treatment effect heterogeneity. Randomization of treatment (or unconfoundedness after appropriate modeling of confounders in observational settings) supports causal inference with respect to treatment in a subgroup (some covariate status). This is subtly, but importantly, different from the assumption that subgroup membership causes or explains a change in treatment effect. For instance, subgroup membership might just be correlated with an unobserved underlying cause of treatment effect heterogeneity. Such implicit inversing of the causal relation underlies the exploration of subgroups as identified by individualized treatment effect prediction to i) inform treatment decisions, or ii) 'explain' treatment effect heterogeneity. However, such conclusions require randomization or unconfoundedness of subgroup membership. Care should be taken to emphasize that the estimated treatment effect describes the causal effect of treatment within a given subgroup, and not necessarily the other way around (\textit{i.e.} it does not warrant the interpretation that the characteristics defining a subgroup cause differential treatment effect). Therefore, while predicted treatment effect heterogeneity may provide interesting hypotheses about its causal structure, it does not provide answers without further thinking about, and analysis of, the causal pathways involved.

\section{Discussion} \label{sec:discussion}

We have provided an overview of the process of individualized treatment effect prediction in the context of a randomized trial with a binary endpoint. To that effect, we have described the integration of key elements from the fields of causal inference and clinical prediction research. These methods can be used to expand on the mainstay analysis of randomized trials, and may help to uncover between-subject heterogeneity in terms of predicted outcome risk and treatment effect. 

With respect to causal inference, we focused on the causal nature of the question of interest and a clear definition of individualized treatment effect based on the potential outcomes framework. From there, we explained the necessary assumptions to identify the individualized treatment effect based on the observed data. While such effects can in principle be estimated nonparametrically, further modeling is beneficial and allows straightforward comparison of treatment effect conditional on many covariates. Even though the prediction problem itself could be solved without any reference to causal inference, going through the motions increases clarity of the research question and gains understanding of the requirements for a causal interpretation of the final model.

With respect to prediction modeling, we focused on the need for a parsimonious model with validity across the risk scale (log odds), while maintaining an interpretable scale for the final result (the risk difference scale). Specification of models for a homogeneous treatment effect (constant relative effect) and differential or heterogeneous treatment effect were described in detail. Subsequently, the relation between prior knowledge and data-driven methodology was examined, revealing the need for both. In line with general sample size guidance when developing a multivariable prediction model \cite{riley_minimum_2019}, sufficient sample size was important for accurate individualized treatment effects predictions and model stability. 

While all of the required ingredients for individualized treatment effect prediction are well-known, their successful combination constitutes a challenging problem that is on the boundary of what can be observed in empirical data. Our simulation study, with a known data generating mechanism, provided clear insight into methods that are able to pick up heterogeneity in sufficiently large samples, while limiting the amount of overfitting in absence of heterogeneity. While very informative, actual analysis of observed data will have to rely on dichotomous outcomes from subjects observed under a single treatment condition that are only incompletely matched across treatment groups. The best way to evaluate the performance of individualized treatment effect prediction models is an open question. We described a bootstrap-based internal validation approach that decreases the risk of overfitting. A very recent contribution to the literature on potential outcome prediction and individualized treatment effect describes a very similar split sample approach \cite{nguyen_counterfactual_2020}. Also, a novel type of c-index has been suggested to measure discriminative performance of individualized treatment effect predictions \cite{van_klaveren_proposed_2018}. Nguyen et al. provide some cautionary notes on its interpretation and estimated standard error in their appendix \cite{nguyen_counterfactual_2020}.

The prediction of individualized or personalized treatment effect is an active field of research. Recent broad overviews on predictive approaches towards heterogeneity of treatment effect are available elsewhere and include a comprehensive overview of applied papers \cite{kent_predictive_2020, rekkas_predictive_2020}. Related work approaches the problem from the missing data perspective \cite{lamont_identification_2018, buuren_flexible_2018}. Also, work has been done on individualized treatment effect prediction for optimal treatment selection \cite{cai_analysis_2011, kang_combining_2014} and selection of patients for future studies \cite{zhao_effectively_2013, li_predictive_2016}. All in all, the literature on personalized medicine approaches that use prediction modeling is vast and too extensive to cite here. What we add is a clear, principled and from the ground-up overview that integrates prediction modeling with causal inference and accentuates the importance of study design features.  

To limit the scope, we did not venture into the incorporation of post-randomization measurements, dropout and selection bias, and observational data. Such topics require careful attention to the exchangeability assumption, which is no longer fulfilled by the study design and needs further assumptions and careful modeling with respect to all possible confounders. A recent scoping review provides an overview of the literature with respect to methods for causal prediction that extend to observational data \cite{lin_scoping_2021}. Also, where we have focused on intention to treat estimates of point exposure treatment, different settings and questions require further thought on the relevant definition of the estimand \cite{van_geloven_prediction_2020}. As a second limitation, we provided a small simulation study covering a limited number of settings that was designed for illustrative purposes. The setup was such that development and test sets were generated from the same data generating mechanism. In practice, there may be differences between these two settings that are not captured by the models, and the uncertainty that accompanies these unknowns may overshadow relatively small gains realized by more complex models \cite{hand_classifier_2006}. 

More general limitations pertain to the typical randomized trial design that provides the data to be used for individualized treatment effect prediction. Other designs, such as $N \times 1$ trials and cross-over designs may provide more direct within-person comparability, and thereby also provide information on the stability of treatment response for an individual \cite{senn_statistical_2018}. However, these designs are infeasible for many conditions and have their own set of challenges \cite{senn_cross-over_2002}. More importantly, randomized trials are typically designed to be of sufficient sample size to reveal an anticipated average relative treatment effect. Therefore, randomized trials are not designed for complex prediction modeling. Hence, if we want to walk down the avenue of individualized treatment effect modelling, we will either have to design trials with this purpose in mind, or have to find more creative ways to amplify our data. This could include the analysis of individual patient data from multiple randomized trials, or even the use of non-randomized studies for the estimation of outcome risk under a control condition \cite{lin_scoping_2021}. Besides clear opportunities, such approaches also bring about many new challenges. For instance, typical challenges that occur in clustered data settings (\textit{e.g.} between-study heterogeneity) have been comprehensively illustrated in a recent tutorial on the examination of heterogeneous (relative) treatment effect in patient-level data from multiple randomized trials \cite{riley_individual_2020}. The implications of such challenges in the context of causal prediction research require further study.

\subsubsection*{Concluding remarks}
We hope that our overview on the basics underlying individualized treatment effect prediction in binary endpoint settings is a useful guide and starting point for statisticians interested in this area. The successful implementation of individualized treatment effect prediction requires careful thought on the exact nature of the question and estimand(s) of interest, the causal and modeling assumptions relied on, and the ever-present bias-variance trade-off that requires even greater care than usual when working with potential outcomes. Sample size considerations are important in all areas of research and there is increasing awareness on the need for larger sample sizes when developing prediction models and examining treatment-covariates interaction. Future work is needed on the validation of models for predicted individualized treatment effect, their role in uncovering sources of heterogeneity, and ways to account for the clustered nature of many data sets. Also, beyond the frequentist framework, the basis for a fully Bayesian approach has long been recognized \cite{simon_bayesian_2002} and could combine the advantage of penalization with a more thorough view on the posterior distribution of the model parameters. A summary of key recommendations and findings is provided in Box \ref{box:keyPoints}.

\begin{minipage}{\linewidth} 
\begin{theo}[Key points and recommendations]{box:keyPoints} 
    \begin{itemize} 
        \item It is important to clearly define the individualized treatment effect of interest and to be aware of the identifiability assumptions underlying its causal interpretation. 
        \item Analogous to causal inference with respect to average treatment effect, randomization of treatment greatly facilitates causal inference with respect to predicted individualized treatment effects.
        \item Logistic regression provides a parsimonious model to predict absolute individualized treatment effect (\textit{i.e.}, treatment effect on the risk difference scale) in new patients. Even in absence of treatment-covariate interaction (\textit{i.e.}, homogeneous patient-level treatment effect on odds ratio scale), a logistic model accounting for individual patient characteristics (prognostic factors) can lead to meaningful differentiation in terms of absolute treatment effect.
        \item The inclusion of treatment-covariate interactions to model heterogeneous treatment effect (\textit{i.e.}, patient-level heterogeneity in the treatment effect on the odds ratio scale) should preferably be based on biological and clinical rationale and be informed by statistical evidence from prior studies.
        \item Sample size and penalized estimation are of key importance for accurate individualized treatment effect prediction; penalization alone does not guarantee accurate predictions in new individuals when sample size is insufficient with respect to model complexity. Existing guidelines on clinical prediction modeling provide a lower bound for the sample size needed in case of individualized treatment effect prediction \cite{riley_minimum_2019, van_smeden_sample_2018, riley_calculating_2020}.
        \item In practice, bootstrap internal validation of likelihood-based measures of overall fit (\textit{e.g.} $\textnormal{R}^2$, AIC), mean squared prediction error (\textit{e.g.} Brier score),  and aggregate observed versus expected measures of treatment effect variability (as in Section \ref{sec:ist}) help to choose amongst competing models. It is recommended to include a homogeneous treatment effect model as a starting point (reference model) and to consider more complex models based on biological, clinical and statistical evidence.
        \item Future work is needed to further delineate best practices for the evaluation of individualized treatment effect predictions and models, hence also improving model comparison and validation procedures. 
        \item Purely explorative indications of heterogeneous treatment effect provide an interesting starting point for further research (e.g. into the causal structure of the heterogeneity) and require external validation. 
        \item This tutorial handles causal prediction of treatment effect on a binary outcome, conditional on individual level patient characteristics. This should not be confused with a causal interpretation of the effect of individual level patient characteristics on the effect of treatment. 
    \end{itemize}
\end{theo}
\end{minipage}

\section*{Data Availability Statement}
Data for the International Stroke Trial applied example are publicly available \cite{international_stroke_trial_collaborative_group_international_2011}. Data for the otitis media applied example not available for sharing since they contain privacy sensitive data according to the General Data Protection Regulation. \texttt{R} scripts to perform the simulation study, including data generation and analysis, are available for sharing.

\section*{Acknowledgements}
We thank the researchers involved in the original otitis media \cite{le_saux_randomized_2005} and stroke trials \cite{international_stroke_trial_collaborative_group_international_1997} for use of their data. 

J Hoogland, M Belias, J in 't Hout, MM Rovers, TPA Debray and JB Reitsma acknowledge financial support from the Netherlands Organisation for Health Research and Development (grant 91215058). TPA Debray also acknowledges financial support from the Netherlands Organisation for Health Research and Development (grant 91617050)

\bibliographystyle{ieeetr}
\bibliography{R2}

\renewcommand\thefigure{\thesection.\arabic{figure}}    
\setcounter{figure}{0} 

%%TC:ignore

\appendix
\numberwithin{equation}{section} 

\section*{Supporting Material}

The online supporting material consists of further information on separate prediction modeling per potential outcome (Part \ref{app:perArm}), the generation of data for a given outcome prevalence (Part \ref{app:eventRate}), and calibration results for the simulation study (Part \ref{app:deltaihat_calibration}).

\newpage

\section{Separate modeling of each potential outcome} \label{app:perArm}

This online supporting material describes the equivalence between a special case of the heterogeneous treatment effect model and models fitted separately in each arm of the trial (section \ref{app:equivalence}), the loss of this equivalence when introducing penalization (section \ref{app:pML}), and simulation results comparing treatment-interaction models with models fitted per treatment arm (section \ref{app:perarmsims}).

\subsection{Equivalent model specifications} \label{app:equivalence}
A logistic heterogeneous treatment effect model as introduced in section \ref{sec:hte} includes both main covariate effects and treatment-covariate interactions. When all covariates (or expansions thereof) in such a model interact with treatment, an exactly equivalent set of 2 models can be specified within the control group and the treated group separately. For instance, a heterogeneous treatment effect model of the form
\begin{align} \label{eq:interaction1}
\begin{split}
\textnormal{logit}(P(Y_{i} = 1 | A=a_i, \bm{X}=\bm{x}_{i})) = 
    \beta_0 + \beta_t a_i + \bm{\beta_m}^\top \bm{x}_{i}+ 
    \bm{\beta_z}^\top \bm{x}_{i} a_i
\end{split}
\end{align}
has a corresponding set of within-treatment group models given by
\begin{align}  \label{eq:perarm1}
\begin{split}
\textnormal{logit}(P(Y_{i} = 1 | A=0, \bm{X}=\bm{x}_{i})) &= 
    \beta_0 + \bm{\beta_m}^\top \bm{x}_{i}\\
\textnormal{logit}(P(Y_{i} = 1 | A=1, \bm{X}=\bm{x}_{i})) &= 
    (\beta_0 + \beta_t) + (\bm{\beta_m} + \bm{\beta_z})^\top \bm{x}_{i}
\end{split}
\end{align}
Note that these two models are separate models for the potential outcomes of interest (\textit{i.e.} $P(Y^{a=0}=1|\bm{X}=\bm{x_i})$ and $P(Y^{a=1}=1|\bm{X}=\bm{x_i})$ respectively). The other way around, starting from two separate models for both potential outcomes as fitted within each treatment group separately, the models 
\begin{align} \label{eq:perarm2}
\begin{split}
\textnormal{logit}(P(Y_{i} = 1 | A=0, \bm{X}=\bm{x}_{i})) &= 
    \beta_{00} + \bm{\beta_{m0}}^\top \bm{x}_{i} \\
\textnormal{logit}(P(Y_{i} = 1 | A=1, \bm{X}=\bm{x}_{i})) &= 
    \beta_{01}+ \bm{\beta_{m1}}^\top \bm{z}_{i} \\
\end{split}
\end{align}
are equivalent to 
\begin{align}  \label{eq:interaction2}
\begin{split}
\textnormal{logit}(P(Y_{i} = 1 | A=a_i, \bm{X}=\bm{x}_{i})) &= 
    \beta_{00} + (\beta_{01}-\beta_{00}) a_i + \bm{\beta_{m0}}^\top \bm{x}_{i} + (\bm{\beta_{m1}} - \bm{\beta_{m0}})^\top \bm{x}_{i} a_i
\end{split}
\end{align}
The equivalence between these model specifications holds for the maximum likelihood estimates of the $\bm{\beta}$ parameter vector, but no longer holds when introducing a penalty into the estimation process. 

\subsection{Penalized maximum likelihood} \label{app:pML}

In case of penalized maximum likelihood, estimates for the separate within treatment-group models will no longer be equivalent to those from a full sample interaction model. For instance, let us consider the case of a ridge or lasso penalty (\textit{i.e.} $\lambda \frac{1}{2} \| \bm{\beta}\|_2^2$ or $\lambda \| \bm{\beta}\|_1$ respectively  \cite{friedman_regularization_2010}). First, each of the models will have its own estimate of $\lambda$, allowing for differences between the within-treatment group models. Second, intercepts are not penalized, and in case of separate models (\textit{e.g.} equation \eqref{eq:perarm1} and \eqref{eq:perarm2}), the main treatment effect is retrieved as the different between the two model intercepts (equation \eqref{eq:interaction2}). Hence, the main treatment effect is penalized by default in the full sample interaction model and is not penalized when using two separate models. Third, in case of ridge regression, the degree of penalization depends on the size of the model coefficients, with larger coefficients being penalized more heavily (due to the square in the penalty term). This is of importance for the treatment-covariate interaction models as specified in equation \eqref{eq:interaction1} and \eqref{eq:interaction2}, since the expression of the covariate effects under treatment and control conditions is not symmetric in that case (\textit{i.e.} with $\bm{\beta_m}$ in equation \eqref{app:equivalence} reflecting covariate effects under the control condition and $\bm{\beta_z}$ reflecting changes from $\bm{\beta_m}$ under the treated condition).  

\subsection{Simulation results} \label{app:perarmsims}

\begin{figure}[!htb]
\centering
\includegraphics[width=.9\linewidth]{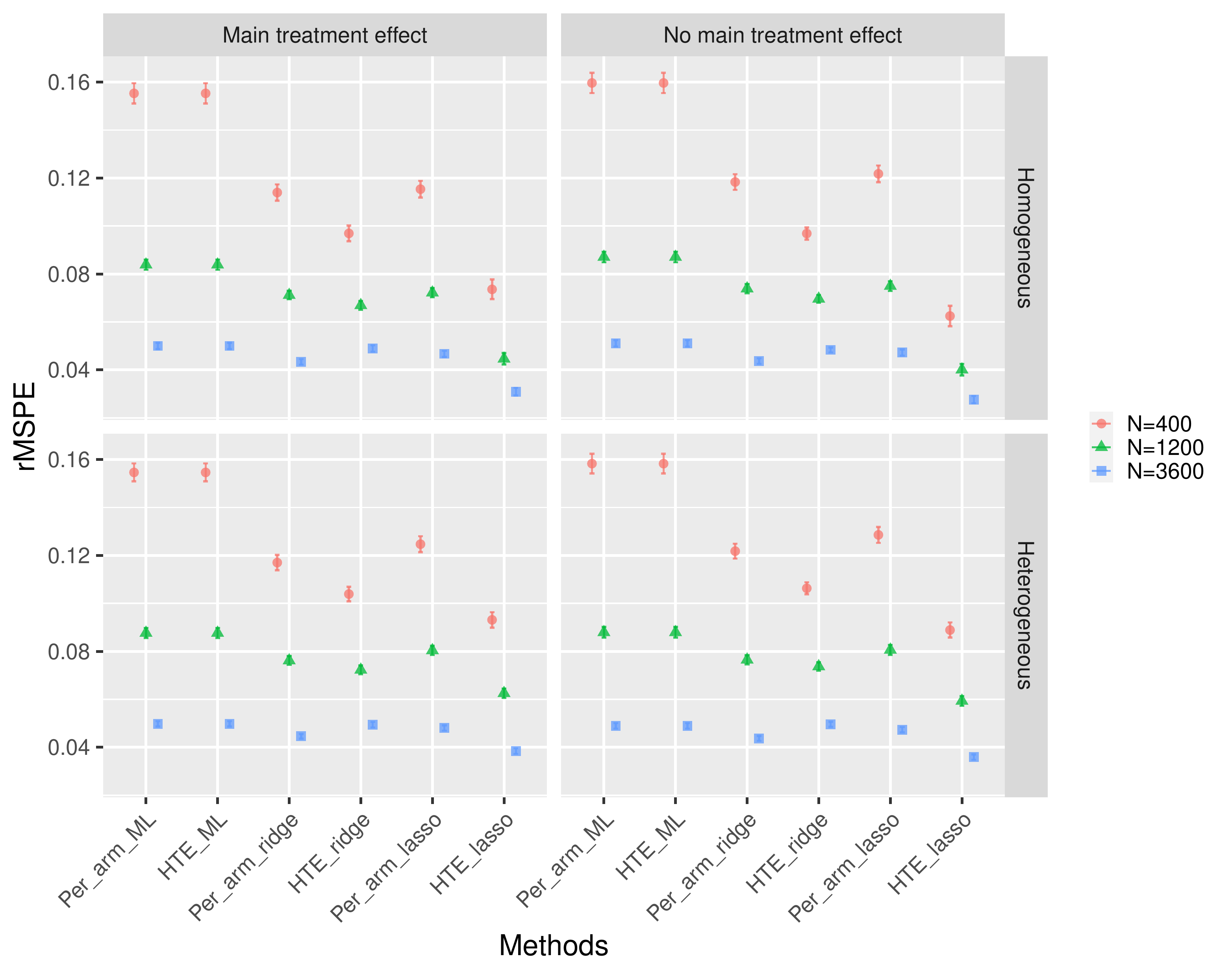}
\caption{Simulation study results: average root mean squared prediction error of the predicted treatment effects (over 250 simulations) with $\pm 2$ SE error bars for all simulation settings. Note that the standard errors are often so small that they are obscured by the mean estimates. Abbreviations for the methods are: HTE (heterogeneous treatment effect model), ML (maximum likelihood)}
\label{fig:perArm}
\end{figure}

The simulation settings were exactly the same as in the main text. Full sample HTE models including all treatment-covariate interactions were compared to within-treatment group models including only main effects of the covariates. Models were estimated by means of maximum likelihood, ridge regression, and lasso regression. Figure \ref{fig:perArm} shows the simulation study results with respect to root mean squared prediction error (rMPSE) of the individualized treatment effects. In case of maximum likelihood estimation, the results are of course exactly the same for the different model specifications and are only shown in twofold as a reminder. Lasso treatment-covariate interaction models performed best across all settings. Also, the rMSPE of predicted individualized treatment effects based on ridge treatment-covariate interaction models was generally better than the prediction error for ridge models fitted separately per arm. One exception was ridge regression in large sample size ($N=3600$), where the per arm models resulted in a better rMSPE. In our simulation settings, which all had variability in coefficient size in the data generating mechanism, the ridge penalty induced clear overshrinkage on large coefficients for all models. This is to be expected due to the square in the penalty and happened in both within-treatment group models and treatment-interaction models. However, in case of treatment-interaction modeling, underfitting of large main effects led to overfitting of the corresponding treatment-covariate interactions \footnote{Note that the square in the ridge penalty means that large estimated coefficients have a larger contribution to the penalty, and are thus more heavily penalized towards zero. An inadvertent characteristic of the treatment-interaction model in case of ridge regression is that the cost of increasing a large main effect parameter (\textit{i.e.} in this context an increase in the effect of the covariate under the control condition), is larger than the cost of the same increase in the smaller corresponding treatment-covariate interaction (\textit{i.e.} the same increase in the effect of the covariate but now under the treatment condition). As a numerical example, assume a main effect coefficient is actually 1 and the corresponding treatment-covariate interaction coefficient is actually 0.5. Shrinking 1 to 0.9 reduces $\| \bm{\beta}\|_2^2$ by 0.19, and overfitting 0.5 by the same amount increases $\| \bm{\beta}\|_2^2$ by only 0.11}.  While this happened in all settings and thus across all sample sizes, we hypothesize that the negative effect of this bias on the predictions $\bm{\delta}$ was offset by more accurate estimation of $\lambda$ in the full sample treatment-interaction models, except in large sample size settings. Therefore, different model specifications that affect to expected size of the estimated coefficients require careful thought in presence of a ridge penalty. These issues do no affect the lasso penalty. In case of lasso regression, the benefit of having a larger sample size to estimate the penalty parameter $\lambda$ (\textit{i.e.} as in the treatment-interaction model) led to better performance in all simulation settings. 

\newpage

\section{Simulating data for a given outcome prevalence} \label{app:eventRate}

The goal was to simulate data with a prespecified outcome prevalence for the control group. The model underlying the simulations was given in equation \eqref{eq:logistic_hte} and is restated here for ease of reference:

\begin{align} 
\begin{split}
\textnormal{logit}(P(Y_{i} = 1 | A=a_i, \bm{X}=\bm{x}_{i})) = 
    \beta_0 + \beta_t a_i + \bm{\beta_m}^\top \bm{x}_{i} + 
    \bm{\beta_z}^\top \bm{z}_{i} a_i
\end{split}
\end{align}

For any given treatment condition, this reduces to 

\begin{align} 
\begin{split}
\textnormal{logit}(P(Y_{i} = 1 | A=a_i, \bm{X}=\bm{x}_{i})) = 
    \beta_{0*} + \bm{\beta_*}^\top \bm{x}_{i}
\end{split}
\end{align}

where $\beta_{0*}$ combines $\beta_0$ and $\beta_t$ and $\bm{\beta_{*}}$ combines $\bm{\beta_{m}}$ and $\bm{\beta_{z}}$. Therefore, conditional on treatment condition, the log odds of an event is a linear combination of just the $p$ covariates. Since these had a standard normal distribution by design, their linear combination is also normal with mean equal to $\beta_{0*}$ and variance equal to 

\begin{equation}
\textnormal{Var}(\beta_{0*} + \bm{\beta_{*}}^\top \bm{x}_{i}) = \bm{\beta \Sigma \beta^\top} = \sigma^2
\end{equation}

where $\beta = \{\beta_{0*},\bm{\beta_{*}}\}$ and $\bm{\Sigma}$ is the covariance matrix of the covariates.

Then using 

\begin{equation}
\textnormal{Pr}(Y=1) = \frac{1}{1 + e^{-\beta_{0*} - \sigma Z}}
\end{equation}

where $Z$ is a standard normally distribute random variable, the outcome prevalence or expected probability of $\textnormal{Pr}(Y=1)$ equals

\begin{align} 
\begin{split}
\mathbb{E}(\textnormal{Pr}(Y=1|\bm{X}))&= \int_{-\infty}^{+\infty} \left( \frac{1}{\sqrt[]{2\pi}} e^{-z^2 / 2} 
												\frac{1}{1 + e^{-\beta_{0*} - \sigma Z}} \right) dz \\
									   &= \frac{1}{\sqrt[]{2\pi}} \int_{-\infty}^{+\infty} 
                                       			\left( \frac{e^{-z^2 / 2}}{1 + e^{-\beta_{0*} - \sigma z}} \right) dz
\end{split}
\end{align}

Since $ \sigma = \sqrt[]{\bm{\beta \Sigma \beta^\top}}$ only depends on known simulation parameters, the equation can be solved numerically for $\beta_{0*}$ to get the desired outcome prevalence in a given treatment group.

\section{Simulation study calibration results} \label{app:deltaihat_calibration}

The \texttt{CalibrationFigures.pdf} file contains calibration plots for $\hat\delta(\bm{x}_i)$, as predicted by each method, versus the true $\delta(\bm{x}_i)$. Simulation settings with a main treatment effect are denoted as $\beta_t < 0$, settings with a homogeneous treatment effect are denoted as HOM, and settings with a heterogeneous treatment effect as HTE. Each individual plot shows the ideal diagonal in red (with an exception of the absolute null settings where the ideal is $\hat\delta(\bm{x}_i) \equiv 0$). Each black calibration line is the result of a single simulation run and connects the mean predicted $\hat\delta(\bm{x}_i)$ and mean $\delta(\bm{x}_i)$ in 20 equal-size quantile groups of $\hat\delta(\bm{x}_i)$. The histograms on the x-axis gives an indication of the density of quantile groups over all simulation (the groups vary due to sampling variability).

\clearpage

%%TC:endignore

\end{document}